\documentclass[12pt]{article}
\def\theequation{\arabic{section}.\arabic{equation}}
\usepackage{amssymb}

\newcounter{rown}

\begin{document}
\renewcommand{\thefootnote}{\fnsymbol{footnote}}
\renewcommand{\theequation}{\thesection.\arabic{equation}}
\title{Constrained
generalized supersymmetries and superparticles with tensorial central
charges.\\ A classification.}
\author{Zhanna Kuznetsova${}^{a}$\thanks{{\em e-mail: zhanna@cbpf.br}}~ and Francesco Toppan${}^{b}$\thanks{{\em e-mail: toppan@cbpf.br}}
\\ \\
${}^a${\it Dep. de F\'{\i}sica, Universidade Estadual de Londrina,}\\{\it Caixa Postal 6001, Londrina (PR), Brazil}\\ ${}^b${\it CBPF, Rua Dr.}
{\it Xavier Sigaud 150,}
 \\ {\it cep 22290-180 Rio de Janeiro (RJ), Brazil}}
\maketitle
\begin{abstract}

We classify the admissible types of constraint (hermitian, holomorphic, with reality
conditions on the bosonic sectors, etc.) for generalized supersymmetries in the presence of
complex spinors. We further point out which constrained generalized supersymmetries
admit a dual formulation.\par
For both real and complex spinors generalized supersymmetries are constructed and classified as dimensional reductions of supersymmetries
from {\em oxidized} space-times (i.e. the maximal space-times associated to $n$-component Clifford 
irreps).\par
We apply these results to sistematically construct a class of  models describing superparticles in presence of bosonic tensorial central charges,
deriving the consistency conditions for the existence of the action,
as well as the constrained equations of motion.
Examples of these models (which, in their twistorial formulation, describe towers of higher-spin particles) were first introduced by Rudychev and Sezgin (for real spinors) and later by Bandos and Lukierski (for complex spinors). 
 
\end{abstract}
\vfill 

\section{Introduction.}

This work addresses two related problems concerning generalized supersymmetries. A mathematical
classification of generalized supersymmetries and of the consistent types of constraint which 
can be applied on them is here furnished. Moreover, an example of application of these results is provided, the classification being used to construct and get information about a class of
dynamical systems which goes under the name of ``superparticles with tensorial central charges".\par
More specifically, on the mathematical side, the main results of the present paper can be summarized as follows. Generalized supersymmetries, for both real and complex spinors,
are classified and explicitly constructed. Specific formulas are given in order to 
express generalized supersymmetries in non-maximal spacetimes as dimensional reduction 
of the corresponding oxidized forms of generalized supersymmetry (i.e., associated with the given maximal space-time carrying the same type of fundamental spinors). For {\em complex} generalized
supersymmetries it is here proven that a consistent set of constraints (holomorphic, hermitian,
reality condition on the bosonic sector, etc.) and of their combinations can be implemented.
A whole subclass of constrained generalized supersymmetries can be produced accordingly.
It is worth mentioning that several such constrained generalized supersymmetries admit
a dual formulation. They can in fact be presented in different, but equivalent forms.
Their dual formulations are explicitly pointed out.\par
On physical side, we addressed the question on how to concretely implement the above
construction and classification in application to dynamical systems. We investigated
the superparticles with tensorial central charges. Specific examples of these models,
for real spinors, have been explicitly constructed at first by Rudychev and Sezgin \cite{rs}. Later
Bandos and Lukierski \cite{bl} analyzed the Minkowskian case with complex spinors. It was further
proven \cite{bls} that, in its twistorial formulation, this system describes towers of higher
spin particles.\par
We introduced the main algebraic ingredients to formulate the action. We checked the consistency condition for the presence of, e.g., a mass term. We derived, using
our previous mathematical results, the number of bosonic degrees
of freedom and of lagrangian multipliers entering the action. In the complex case (i.e. for
the formulation in terms of complex spinors) we pointed out several distinct possibilities for
constructing the action (corresponding to a given choice of the metric for spinors).
The constrained models, associated to the constrained generalized supersymmetries, have been
constructed and classified in terms of their main properties.\par
Some words should be spent to explain the main motivations of the present work. The mathematical classificatory aspect is a continuation of previous works \cite{{crt1},{top}},
on the classification of spinors and generalized supersymmetries (this line of research,
based on several mathematical works \cite{{abs},{por},{oku}}, is currently under intense investigation,
see also the references \cite{{dflv},{acdp}}). In particular in \cite{top} the notion of constrained holomorphic and
hermitian supersymmetry has been introduced (a physical application of a holomorphic constrained generalized supersymmetry can be found in \cite{lt3},
where the notion of Euclidean $M$ algebra has been introduced). The present work extends this analysis, 
proving the existence of whole new subclasses of constrained generalized supersymmetries.\par
The physical interest in generalized supersymmetries is of course mainly motivated by the $M$-theory related
investigations \cite{{agit},{tow}}. The first example of a generalized supersymmetry going beyond the 
Haag- \L opusza\'{n}ski-Sohnius scheme \cite{hls} was given in \cite{daf}. It is nowadays recognized
that generalized supersymmetries are associated with the dynamics of extended objects like branes \cite{{gs},{ste}}.\par
Even if the $M$-algebra is essentially unique and based on $11$-dimensional Minkowskian real spinors, it is known that a whole web of dualities
relate different theories arising as dimensional reductions from a given oxidized theory.
It is worth pointing out that several dynamical systems presenting the same symmetry algebra
(or some of its generalizations) expected by the $M$-theory \cite{west} can be constructed.
An example of such class of models is given by the higher-dimensional Chern-Simons
supergravities, see \cite{htz}.\par
Somehow the simplest scenario where the algebraic setup of generalized supersymmetries
(and of their consistent constraints) can be implemented in a dynamical setting is given by the
superparticles in presence of bosonic tensorial central charges. Extra ingredients
(which will be conveniently specified in the following) beyond the purely algebraic data entering the generalized supersymmetries, have to be introduced in order to construct such 
models. It is worth noticing, however, that a good deal of information about these theories 
can be recovered just by using the mathematical data of the present classification.\par
It must be said that the interest in these models is not merely academical. Indeed, it has been proven that they are linked with the Fronsdal's proposal \cite{fro} of introducing bosonic tensorial coordinates in order to deal with higher spin theories. It was shown in \cite{bls} that, explicitly solving the constraints arising from the lagrange multipliers in terms of twistors,
allows to recover a massless higher spin field theory, as the one arising in the tensionless
limit of the superstrings (see \cite{sor} and references therein). Several groups are presently investigating these theories as a possible viable approach to the problem
of formulating a consistent string field theory which should, hopefully, be regarded as a
spontaneous breaking of the conformally invariant theory of the massless higher spin particles.\par
Beyond superparticles, other type of models like strings with tensorial central charges can be
constructed and physically motivated \cite{zlu}. Even if we do not address these types
of theories here, nevertheless some of the results here obtained can be applied to investigate
this class of models as well. \par
The scheme of this paper is as follows. The next two following sections contain no new result, but present in compact form the needed mathematical ingredients and conventions used throughout the paper. Section {\bf 4} is devoted to explain the dimensional reduction of
Clifford irreps from oxidized spacetimes. Section {\bf 5} introduces the superparticle models with tensorial central charges. The sections {\bf 6} and {\bf 7} are devoted to the mathematical
aspects of {\em real} generalized supersymmetries and to the classification of
superparticles with real spinors. Complex generalized supersymmetries are discussed 
in sections {\bf 8} and {\bf 9}. The classification of the constrained generalized supersymmetries and of their duality relations is given in section {\bf 10}. Sections
{\bf 11} and {\bf 12} are devoted to the construction and classification of complex
superparticles with tensorial central charges. 
Finally, in the Conclusions, some further discussions are made on the possibile
applications of the present results.

\section{Clifford irreps and fundamental spinors.}

This section is devoted to a quick review, following \cite{crt1} and \cite{top}, of the fundamental ingredients and conventions entering the classification of generalized supersymmetries. It is intended to make this paper self-consistent. More detailed explanations can be found in the cited references.\par
The Clifford algebra
\begin{eqnarray}
\Gamma^\mu\Gamma^\nu+\Gamma^\nu\Gamma^\mu &=& 2\eta^{\mu\nu},
\label{cliff}
\end{eqnarray}
with $\eta^{\mu\nu}$ being a diagonal matrix of $(p,q)$ signature
(i.e. $p$ positive, $+1$, and $q$ negative, $-1$, diagonal
entries) admits irreps which are classified according to the 
most general $S$ matrix commuting with all the $\Gamma$'s ($\relax
[S,\Gamma^\mu ] =0$ for all $\mu$). The most general $S$ can be a multiple of the identity
(real ${\bf R}$ case), ``almost complex" (the ${\bf C}$ case) or the linear combination of four matrices closing the quaternionic algebra (the ${\bf H}$ case).\par
We recall that the spinors carry a representation of the $so(p,q)$ algebra of commutators
$\Sigma_{\mu\nu}= [\Gamma_\mu, \Gamma_\nu ]$. In the space-times where the Gamma-matrices can 
be chosen to be of block-antidiagonal form, it is possible to introduce the notion of ``generalized" Weyl-projected spinors, whose number of components is half of the size of the corresponding Gamma matrices \cite{crt1}. 
Spinors are called ``fundamental" if their representation of the generalized Lorentz group 
admits minimal number of real components in association with the maximal, compatible, allowed division-algebra structure.\par
A useful table presents the
comparison between division-algebra properties of Clifford irreps (${\bf\Gamma}$) and fundamental spinors (${\bf\Psi}$)
in different space-times parametrized by $\rho= p-q\quad mod\quad 8$.
We have
{ {{\begin{eqnarray}&\label{gammapsi}
\begin{tabular}{|c|c|c|}\hline
$\rho$&${ \Gamma}$ &${ \Psi}$
\\ \hline

$0$&${\bf R}$&${\bf R}$\\ \hline

$1$&${\bf R} $&${\bf R}$\\ \hline

$2$&${\bf R}$&${\bf C}$\\ \hline

 $3$&${\bf C}
$&${\bf H}$\\ \hline

$4$&${\bf H}$&${\bf H}$\\ \hline

 $5$&${\bf H}
$&${\bf H}$\\ \hline

$6$&${\bf H}$&${\bf C}$\\ \hline

 $7$&${\bf C}
$&${\bf R}$\\ \hline

\end{tabular}&\end{eqnarray}}} }
For $\rho=2,3$, the fundamental spinors can accommodate a larger
division-algebra structure than the corresponding Clifford irreps. Conversely, for $\rho= 6,7$, the Clifford irreps
accommodate a larger division-algebra structure than the corresponding spinors.\par
Throughout the text the notation ``$(p,q)_\Gamma$" will be used to denote the Clifford irreps,
while ``$(p,q)_\Psi$" will be employed to denote the Clifford representations which,
under Weyl projection, generate fundamental spinors (the symbol ``$(p,q)$" will appear when the
two constructions coincide).\par
An important notion which will be extensively used in the following is that of ``maximal Clifford algebra". It corresponds to the maximal space-time (of $(p,q)$ signature) which can accommodate a Clifford irrep realized by Gamma matrices of a given size. Non-maximal Clifford algebras are recovered as
dimensional reductions of some maximal Clifford algebra, deleting a certain number of Gamma matrices which are now considered as external \cite{crt1}. The explicit construction of non-maximal Clifford algebras from their associated maximal Clifford algebras (their {\em oxidized} form) is presented in Section {\bf 4}.\par
Quite a convenient algorithmic procedure to sistematically produce
representatives of maximal Clifford algebras irreps in any allowed space-time was explicitly presented in \cite{crt1} and \cite{top}.

\section{Notion of generalized supersymmetries.}

In this section we will briefly review the necessary notions concerning the generalized
supersymmetries. More detailed explanations are given in references \cite{crt1} and \cite{top}.
\par
At first we need to introduce two matrices, denoted as $A$ and $C$ \cite{kt}, related with,
respectively, the hermitian
conjugation and transposition acting on Gamma matrices.
$A$ plays the role of the time-like $\Gamma^0$ matrix in
the Minkowskian space-time and is used to introduce barred spinors. $C$, on the other hand,
is the charge conjugation matrix. Up to an overall sign, in a generic $(p,q)$ space-time, $A$ and $C$
are given by the products of all the time-like and, respectively, all the symmetric (or antisymmetric) Gamma-matrices (depending on the given space-time there are at most two charge conjugations matrices,
$C_S$, $C_A$, given by the product of all symmetric and all antisymmetric gamma matrices).
For our purposes the importance of $A$ and the charge conjugation matrix $C$ lies on the fact that, in a 
$D$-dimensional space-time ($D=p+q$) spanned by $d\times d$ Gamma matrices, they allow to construct a basis for $d\times d$ (anti)hermitian and (anti)symmetric matrices, respectively.
The
$\left( \begin{array}{c}
  D\\
  k
\end{array}\right)$ antisymmetrized products of $k$ Gamma matrices
$A{\Gamma}^{[\mu_1 \ldots \mu_k]}$ are all hermitian or all antihermitian, depending on the value 
of $k\leq D$. Similarly, the antisymmetrized products $C{\Gamma}^{[\mu_1 \ldots \mu_k]}$ are all 
symmetric or all antisymmetric.\par
A generalized supersymmetry algebra involving $n$-component real spinors $Q_a$ is given by the anticommutators
\begin{eqnarray}\label{Mgen}
    \left\{ Q_a, Q_b \right\} & = & {\cal Z}_{ab},
\end{eqnarray}
where the matrix ${\cal Z} $ appearing in the r.h.s. is the most general $n\times n$
symmetric matrix with total number of $\frac{n(n+1)}{2}$ components. For any given space-time we
can easily compute its associated decomposition
in terms of the antisymmetrized products of $k$-Gamma matrices, namely
\begin{eqnarray}
{\cal Z}_{ab} &=& \sum_k(C\Gamma_{[\mu_1\ldots\mu_k]})_{ab}Z^{[\mu_1\ldots \mu_k]},
\end{eqnarray}
where the values $k$ entering the sum in the r.h.s. are restricted by the symmetry requirement for the 
$a\leftrightarrow b$ exchange
and are specific for the given spacetime. The coefficients $Z^{[\mu_1\ldots \mu_k]}$ are the rank-$k$ abelian tensorial central charges.  \par
In the case of Weyl projected spinors ${\widetilde Q}_a$ the r.h.s. has to be reconstructed with the help of a projection operator which selects the upper left block in a $2\times 2$ block decomposition.
Specifically, if ${\cal Z}$ is a matrix decomposed in $2\times 2$ blocks as
${\cal Z} =\left( \begin{array}{cc}
  {\cal Z}_1&{\cal Z}_2\\
  {\cal Z}_3&{\cal Z}_4
\end{array}\right)$, we can define 
\begin{eqnarray}\label{pweyl}
P({\cal Z}) &\equiv & {\cal Z}_1.
\end{eqnarray} 
The generalized supersymmetry algebra in the Weyl case can be expressed through
\begin{eqnarray}\label{weylproj}
    \left\{ {\widetilde Q}_a, {\widetilde Q}_b \right\} & = & P({\cal Z})_{ab}.
\end{eqnarray}
A complex generalized supersymmetry algebra is expressed in terms of complex spinors $Q_a$ and their complex conjugate ${Q^\ast}_{\dot a}$.
The most general (with a saturated r.h.s.) algebra is in this case given by 
\begin{eqnarray}\label{Mhol}
    \left\{ Q_a, Q_b \right\} =  {\cal P}_{ab}\quad &,& \quad \left\{ {Q^\ast}_{\dot a}, {Q^\ast}_{\dot b} \right\} =  {{\cal P}^\ast}_{\dot{a}\dot{b}},
\end{eqnarray}
together with
\begin{eqnarray}\label{Mher}
\left\{ Q_a, {Q^\ast}_{\dot b} \right\} &=&  {\cal R}_{{a}\dot{b}},
\end{eqnarray}
where the matrix ${\cal P}_{ab}$ (${{\cal P}^\ast}_{\dot{a}\dot{b}}$ is its conjugate and does not contain new degrees of freedom) is symmetric,
while ${\cal R}_{{a}\dot{b}}$ is hermitian.\par
The maximal number of allowed components in the r.h.s. is given, for complex fundamental spinors
with $n$ complex components, by $n(n+1)$ (real) bosonic components entering the symmetric $n\times n$ complex matrix ${\cal P}_{ab}$ 
plus $n^2$  (real) bosonic components entering the hermitian $n\times n$ complex matrix 
${\cal R}_{{a}\dot{b}}$.\par
A Weyl projection similar to (\ref{weylproj}) can be applied for complex spinors as well.

\section{Oxidation and dimensional reductions.}

Maximal Clifford algebras (whose definition has been recalled in Section {\bf 2}) are
encountered if and only if \cite{{oku},{crt1}} the condition  $p-q =1,5\quad mod \quad 8$,
for a $(p,q)$ spacetime, is matched. In this section we explicitly construct, via dimensional
reduction, all the non-maximal Clifford algebras obtained from any given maximal Clifford
algebra (i.e. their {\em oxidized} form\footnote{We recall that in the superstrings/$M$-theory literature, the term ``oxidation" is commonly employed to denote the operation corresponding to the inverse of the dimensional reduction \cite{lpss}.}). We generalize here similar formulas produced in \cite{crt1}
for real and quaternionic spinors, by taking into account the complex case as well. More
specifically, we determine how to construct the Clifford irreps (denoted as $(p,q)_\Gamma$),
as well as the Clifford representations associated with fundamental spinors (denoted throughout this paper as $(p,q)_\Psi$, see the comment at the end of Section ${\bf 2}$) from their associated ({\em oxidized}) maximal Clifford algebras. The construction of $(p,q)_\Gamma$
coincides with that of $(p,q)_\Psi$ only in those cases, listed in table (\ref{gammapsi}), where the division algebra properties of the Clifford irreps match the division-algebra properties of
fundamental spinors. \par
For what concerns the explicit construction of the maximal Clifford algebras, it can be carried
out iteratively with the two algorithms presented in \cite{crt1}.\par
As it is clear from table (\ref{gammapsi}) the 
\begin{eqnarray}
p-q &=& 1\quad mod \quad 8
\end{eqnarray}
maximal Clifford algebras are of real (${\bf R})$ type. Similarly, all their dimensionally
reduced Clifford algebras (both the $(p,q)_\Gamma$ Clifford irreps and the $(p,q)_\Psi$
representations carried by fundamental spinors) are of real type.\par
On the other hand the
\begin{eqnarray}
p-q &=& 5\quad mod\quad 8
\end{eqnarray}
maximal Clifford algebras are of quaternionic (${\bf H})$ type. They originate quaternionic
and complex non-maximal (dimensionally reduced) Clifford irreps and representations of fundamental spinors. \par
It is worth recalling that none of the maximal Clifford algebras are of block-antidiagonal, ``Weyl-type", form (as a consequence, their associated spinors are not Weyl-projected). The process of deleting a space-like Gamma matrix always
produces a Weyl-type non-maximal Clifford algebra (see \cite{crt1}). \par
Real non-maximal Clifford irreps ($\Gamma$) and fundamental spinor representations ($\Psi$)
as well as their non-maximal complex or quaternionic (for both ${\bf C}$ and ${\bf H}$) equivalent are {\em generically} constructed according to the table below. It is specified 
how the dimensional reduction  has to be carried out from the corresponding associated oxidized maximal Clifford algebra. We have 
{\small
 { {{\begin{eqnarray}&\label{maxclifford}
\begin{tabular}{|l|l|l|}\hline
$$&${1\quad mod\quad 8\quad ({\bf R})}$&${5\quad mod\quad 8 \quad ({\bf H})}$
\\ \hline

$0\quad mod\quad 8$&$\Gamma,\Psi: (p,q)\stackrel{W}{\rightarrow} (p-1,q)$&$$\\ \hline

$4\quad mod \quad 8$&$ $&$\Gamma,\Psi: (p,q)\stackrel{W}{\rightarrow} (p-1,q)$\\ \hline

$2\quad mod\quad 8$&$\Gamma: (p,q)\stackrel{}{\rightarrow} (p,q-1)$&$
\Psi:(p,q)\stackrel{\ast}{\rightarrow} (p-2,q)\stackrel{W}{\rightarrow} (p-3,q)$\\ \hline

 $3\quad mod\quad 8$&$
$&$\begin{tabular}{l}$\Gamma: (p,q)\stackrel{\ast}{\rightarrow} (p-2,q)$\\
$\Psi: (p,q)\stackrel{W}{\rightarrow} (p-2,q)$
\end{tabular}$\\ \hline

$6\quad mod\quad 8$&$$&$
\begin{tabular}{l}$
\Gamma: (p,q){\rightarrow} (p,q-1)$\\$
\Psi: (p,q)\stackrel{\ast}{\rightarrow} (p,q-2)\stackrel{W}{\rightarrow} (p-1,q-2)
$\end{tabular}$\\ \hline

 $7\quad mod\quad 8$&$\Psi: (p,q)\stackrel{W}{\rightarrow} (p-2,q)
$&$\Gamma: (p,q)\stackrel{\ast}{\rightarrow} (p,q-2)$\\ \hline

\end{tabular}&\nonumber\\&&\end{eqnarray}}} }}
Some remarks are in order. The real case is shown in the second column, while both
the complex and the quaternionic cases are recovered from the third column. The arrows denote
which gamma matrices (either space-like or time-like) and how many of them have to be deleted
from the corresponding maximal Clifford algebra. The ``$W$" symbol above an arrow specifies whether the Weyl projection is required in order to produce fundamental spinors.\par
As it is clear from the algorithmic construction of maximal Clifford algebras given in \cite{crt1},
in a $D$-dimensional, $D=p+q$ space-time, $D-3$ Gamma matrices are real, while the three
remaining ones can be expressed as the three imaginary quaternions multiplying a common
real matrix. Deleting two such imaginary quaternionic matrices produces a non-maximal Clifford algebra with a single imaginary Gamma matrix. In this way one can recover a complex,
non-maximal Clifford algebra from a quaternionic maximal Clifford algebra. The so-produced
complex non-maximal Clifford algebra can be represented with half of 
the original size of the quaternionic maximal Clifford algebra
(the ``$\ast$" symbol above an arrow is inserted to remind this fact).
A convenient illustrative example of this feature concerns the imaginary unit $i$ itself;
while the
whole set of three imaginary quaternions requires a presentation either in terms of 
three $2\times 2$ complex matrices or three $4\times 4$ real matrices, the imaginary unit 
can be realized as a single complex number or a $2\times 2$ real matrix.  \par
The table (\ref{maxclifford}) above is ``generic" in the following sense. A careful inspection reveals
that at least three Gamma matrices associated with the imaginary quaternions are required
to carry the construction specified by the ``$\ast$" symbol above an arrow. Therefore, we can only perform the $(p,q)\stackrel{\ast}{\rightarrow} (p-2,q)$ construction under the condition
$p\geq 3$ and, similarly, the $(p,q)\stackrel{\ast}{\rightarrow} (p,q-2)$ 
construction for $q\geq 3$.\par
Very few exceptional cases cannot be recovered from the above table.
Up to $D=11$, only three non-maximal Clifford algebras 
(given by $(0,5)_\Gamma$, $(6,0)_\Psi$ and $(0,7)_\Gamma$) cannot be produced according to 
(\ref{maxclifford}).
Their explicit construction is given below.\par
It is quite convenient to present the full list of reductions of the
non-maximal Clifford irreps and fundamental spinors,  for any $D$-dimensional space-time up
to $D=11$ (their corresponding division algebra character, ${\bf R}$, ${\bf C}$ or ${\bf H}$, is also presented in the last column). We have
{\small
 { {{\begin{eqnarray}&\label{maxcliff}
\begin{tabular}{|l|lcll|}\hline
$D=1$&$(1,0)_{\Gamma,\Psi}$&:&$ M. C. A.$&(${\bf R}$)\\ 
$$&$(0,1)_\Gamma $&:&$ (0,3)\stackrel{\ast}\rightarrow (0,1)$&(${\bf C}$)\\
$$&$(0,1)_\Psi $&:&$ (2,1)\stackrel{W}\rightarrow (0,1)$&(${\bf R}$)\\ \hline

$D=2$&$(2,0)_{\Gamma}$&:&$ (2,1)\rightarrow (2,0)$&(${\bf R}$)\\ 
$$&$(2,0)_\Psi $&:&$ (5,0)\stackrel{\ast}\rightarrow (3,0)\stackrel{W}\rightarrow(2,0)$&(${\bf C}$)\\

$$&$(1,1)_{\Gamma , \Psi} $&:&$ (2,1)\rightarrow (1,1)$&(${\bf R}$)\\

$$&$(0,2)_\Gamma $&:&$ (0,3)\rightarrow (0,2)$&(${\bf H}$)\\
$$&$(0,2)_\Psi $&:&$(1,4)\stackrel{\ast}\rightarrow (1,2)\stackrel{W}\rightarrow(0,2)$&(${\bf C}$)\\ \hline

$D=3$&$(3,0)_{\Gamma}$&:&$ (5,0)\stackrel{\ast}\rightarrow (3,0)$&(${\bf C}$)\\

$$&$(3,0)_\Psi $&:&$(5,0)\stackrel{W}\rightarrow (3,0)$&(${\bf H}$)\\
 
$$&$(2,1)_{\Gamma , \Psi} $&:&$ M.C.A.$&(${\bf R}$)\\

$$&$(1,2)_\Gamma $&:&$ (1,4)\rightarrow (1,2)$&(${\bf C}$)\\

$$&$(1,2)_\Psi $&:&$ (3,2)\stackrel{W}\rightarrow (1,2)$&(${\bf R}$)\\

$$&$(0,3)_{\Gamma , \Psi} $&:&$ M.C.A.$&(${\bf H}$)\\
 \hline

$D=4$&$(4,0)_{\Gamma,\Psi}$&:&$(5,0)\stackrel{W}\rightarrow (4,0) $&(${\bf H}$)\\ 
$$&$(3,1)_\Gamma $&:&$ (3,2)\rightarrow (3,1)$&(${\bf R}$)\\
$$&$(3,1)_\Psi $&:&$(6,1)\stackrel{\ast}\rightarrow (4,1)\stackrel{W}\rightarrow (3,1)$&(${\bf C}$)\\ 
$$&$(2,2)_{\Gamma, \Psi} $&:&$ (3,2)\stackrel{W}\rightarrow (2,2)$&(${\bf R}$)\\
$$&$(1,3)_\Gamma $&:&$ (1,4)\rightarrow (1,3)$&(${\bf H}$)\\ 
$$&$(1,3)_\Psi $&:&$ (2,5)\stackrel{\ast}\rightarrow (2,3)\stackrel{W}\rightarrow (1,3)$&(${\bf C}$)\\
$$&$(0,4)_{\Gamma , \Psi} $&:& $(1,4)\stackrel{W}\rightarrow (0,4)$&(${\bf H}$)\\ 
\hline

$D=5$&$(5,0)_{\Gamma,\Psi}$&:& $M.C.A.$&(${\bf H}$)\\ 
$$&$(4,1)_\Gamma $&:& $(6,1)\stackrel{\ast}\rightarrow (4,1)$&(${\bf C}$)\\
$$&$(4,1)_\Psi $&:&$ (6,1)\stackrel{W}\rightarrow (4,1)$&(${\bf H}$)\\ 
$$&$(3,2)_{\Gamma, \Psi} $&:&$ M.C.A.$&(${\bf R}$)\\
$$&$(2,3)_\Gamma $&:&$ (2,5)\stackrel{\ast}\rightarrow (2,3)$&(${\bf C}$)\\ 
$$&$(2,3)_\Psi $&:&$ (4,3)\stackrel{W}\rightarrow (2,3)$&(${\bf R}$)\\
$$&$(1,4)_{\Gamma , \Psi} $&:&$ M.C.A.$&(${\bf H}$)\\ 
$$&$(0,5)_\Gamma $&:&$ \clubsuit$&(${\bf C}$)\\ 
$$&$(0,5)_\Psi $&:&$ (2,5)\stackrel{W}\rightarrow (0,5)$&(${\bf H}$)\\

\hline

$D=6$&$(6,0)_{\Gamma}$&:&$ (6,1)\rightarrow (6,0)$&(${\bf H}$)\\ 
$$&$(6,0)_\Psi $&:&$ \clubsuit$&(${\bf C}$)\\
$$&$(5,1)_{\Gamma , \Psi} $&:&$ (6,1)\stackrel{W}\rightarrow (5,1)$&(${\bf H}$)\\ 
$$&$(4,2)_{\Gamma} $&:&$ (4,3)\rightarrow (4,2)$&(${\bf R}$)\\
$$&$(4,2)_\Psi $&:&$ (7,2)\stackrel{\ast}\rightarrow (5,2)\stackrel{W}\rightarrow (4,2)$&(${\bf C}$)\\ 
$$&$(3,3)_{\Gamma ,\Psi} $&:&$ (4,3)\stackrel{W}\rightarrow (3,3)$&(${\bf R}$)\\
$$&$(2,4)_{\Gamma} $&:&$ (2,5)\rightarrow (2,4)$&(${\bf H}$)\\ 
$$&$(2,4)_\Psi $&:&$ (3,6)\stackrel{\ast}\rightarrow (3,4)\stackrel{W}\rightarrow (2,4)$&(${\bf C}$)\\ 
$$&$(1,5)_{\Gamma, \Psi} $&:&$ (2,5)\stackrel{W}\rightarrow (1,5)$&(${\bf H}$)\\
$$&$(0,6)_{\Gamma} $&:&$ (0,7)\rightarrow (0,6)$&(${\bf R}$)\\ 
$$&$(0,6)_\Psi $&:&$ (3,6)\stackrel{\ast}\rightarrow (1,6)\stackrel{W}\rightarrow (0,6)$&(${\bf C}$)\\ 
\hline
\end{tabular}&\nonumber\\&&\nonumber\end{eqnarray}}} }}{\small
 { {{\begin{eqnarray}&\label{maxcliff2}
\begin{tabular}{|l|lcll|}\hline
$D=7$&$(7,0)_{\Gamma}$&:&$  \clubsuit$&(${\bf C}$)\\ 
$$&$(7,0)_\Psi $&:&$ (9,0)\stackrel{W}\rightarrow (7,0)$&(${\bf R}$)\\
$$&$(6,1)_{\Gamma , \Psi} $&:&$ M.C.A.$&(${\bf H}$)\\ 
$$&$(5,2)_{\Gamma} $&:&$ (7,2)\stackrel{\ast}\rightarrow (5,2)$&(${\bf C}$)\\
$$&$(5,2)_\Psi $&:&$ (7,2)\stackrel{W}\rightarrow (5,2)$&(${\bf H}$)\\ 
$$&$(4,3)_{\Gamma ,\Psi}$&:&$ M.C.A.$&(${\bf R}$)\\
$$&$(3,4)_{\Gamma} $&:&$ (3,6)\stackrel{\ast}\rightarrow (3,4)$&(${\bf C}$)\\ 
$$&$(3,4)_\Psi $&:&$ (5,4)\stackrel{W}\rightarrow (3,4)$&(${\bf R}$)\\ 
$$&$(2,5)_{\Gamma, \Psi} $&:&$M.C.A.$&(${\bf H}$)\\
$$&$(1,6)_{\Gamma} $&:&$(3,6)\stackrel{\ast}\rightarrow (1,6)$&(${\bf C}$)\\ 
$$&$(1,6)_\Psi $&:&$ (3,6)\stackrel{W}\rightarrow (1,6)$&(${\bf H}$)\\ 
$$&$(0,7)_{\Gamma, \Psi} $&:&$ M.C.A.$&(${\bf R}$)\\ 
\hline
$D=8$&$(8,0)_{\Gamma,\Psi}$&:&$ (9,0)\stackrel{W}\rightarrow (8,0)$&(${\bf R}$)\\ 
$$&$(7,1)_\Gamma $&:&$ (7,2)\rightarrow (7,1)$&(${\bf H}$)\\
$$&$(7,1)_\Psi $&:&$ (8,3)\stackrel{\ast}\rightarrow (8,1)\stackrel{W}\rightarrow (7,1)$&(${\bf C}$)\\ 
$$&$(6,2)_{\Gamma, \Psi} $&:&$ (7,2)\stackrel{W}\rightarrow (6,2)$&(${\bf H}$)\\
$$&$(5,3)_\Gamma $&:&$ (5,4)\rightarrow (5,3)$&(${\bf R}$)\\ 
$$&$(5,3)_\Psi $&:&$ (8,3)\stackrel{\ast}\rightarrow (6,3)\stackrel{W}\rightarrow (5,3)$&(${\bf C}$)\\
$$&$(4,4)_{\Gamma , \Psi} $&:&$ (5,4)\stackrel{W}\rightarrow (4,4)$&(${\bf R}$)\\ 
$$&$(3,5)_\Gamma $&:&$ (3,6)\rightarrow (3,5)$&(${\bf H}$)\\ 
$$&$(3,5)_\Psi $&:&$ (4,7)\stackrel{\ast}\rightarrow (4,5)\stackrel{W}\rightarrow (3,5)$&(${\bf C}$)\\
$$&$(2,6)_{\Gamma, \Psi} $&:&$ (3,6)\stackrel{W}\rightarrow (2,6)$&(${\bf H}$)\\
$$&$(1,7)_\Gamma $&:&$ (1,8)\rightarrow (1,7)$&(${\bf R}$)\\ 
$$&$(1,7)_\Psi $&:&$ (4,7)\stackrel{\ast}\rightarrow (2,7)\stackrel{W}\rightarrow (1,7)$&(${\bf C}$)\\
$$&$(0,8)_{\Gamma , \Psi} $&:&$(1,8)\stackrel{W}\rightarrow (0,8)$&(${\bf R}$)\\ 
\hline
$D=9$&$(9,0)_{\Gamma,\Psi}$&:&$ M.C.A.$&(${\bf R}$)\\ 
$$&$(8,1)_\Gamma $&:&$ (8,3)\stackrel{\ast}\rightarrow (8,1)$&(${\bf C}$)\\
$$&$(8,1)_\Psi $&:&$ (10,1)\stackrel{W}\rightarrow (8,1)$&(${\bf R}$)\\ 
$$&$(7,2)_{\Gamma, \Psi} $&:&$ M.C.A.$&(${\bf H}$)\\
$$&$(6,3)_\Gamma $&:&$ (8,3)\stackrel{\ast}\rightarrow (6,3)$&(${\bf C}$)\\ 
$$&$(6,3)_\Psi $&:&$ (8,3)\stackrel{W}\rightarrow (6,3)$&(${\bf H}$)\\
$$&$(5,4)_{\Gamma , \Psi} $&:&$ M.C.A.$&(${\bf R}$)\\ 
$$&$(4,5)_\Gamma $&:&$ (4,7)\stackrel{\ast}\rightarrow (4,5)$&(${\bf C}$)\\ 
$$&$(4,5)_\Psi $&:&$ (6,5)\stackrel{W}\rightarrow (4,5)$&(${\bf R}$)\\
$$&$(3,6)_{\Gamma, \Psi} $&:&$ M.C.A.$&(${\bf H}$)\\
$$&$(2,7)_\Gamma $&:&$(4,7)\stackrel{\ast}\rightarrow (2,7)$&(${\bf C}$)\\ 
$$&$(2,7)_\Psi $&:&$ (4,7)\stackrel{W}\rightarrow (2,7)$&(${\bf H}$)\\
$$&$(1,8)_{\Gamma , \Psi} $&:&$M.C.A.$&(${\bf R}$)\\ 
$$&$(0,9)_\Gamma $&:&$ (0,11)\stackrel{\ast}\rightarrow (0,9)$&(${\bf C}$)\\ 
$$&$(0,9)_\Psi $&:&$ (2,9)\stackrel{W}\rightarrow (0,9)$&(${\bf R}$)\\
\hline\end{tabular}&\nonumber\\&&\nonumber\end{eqnarray}}} }}
{\small
 { {{\begin{eqnarray}&\label{maxcliff3}
\begin{tabular}{|l|lcll|}\hline
$D=10$&$(10,0)_{\Gamma}$&:&$ (10,1)\rightarrow (10,0)$&(${\bf R}$)\\ 
$$&$(10,0)_\Psi $&:&$ (13,0){\stackrel{\ast}\rightarrow} (11,0){\stackrel{W}\rightarrow} (10,0)$&(${\bf C}$)\\
$$&$(9,1)_{\Gamma , \Psi} $&:&$ (10,1)\stackrel{W}\rightarrow (9,1)$&(${\bf R}$)\\ 
$$&$(8,2)_{\Gamma} $&:&$ (8,3)\rightarrow (8,2)$&(${\bf H}$)\\
$$&$(8,2)_\Psi $&:&$ (9,4){\stackrel{\ast}\rightarrow} (9,2){\stackrel{W}\rightarrow} (8,2)$&(${\bf C}$)\\ 
$$&$(7,3)_{\Gamma ,\Psi}$&:&$ (8,3){\stackrel{W}\rightarrow} (7,3)$&(${\bf H}$)\\
$$&$(6,4)_{\Gamma} $&:&$ (6,5)\rightarrow (6,4)$&(${\bf R}$)\\ 
$$&$(6,4)_\Psi $&:&$ (9,4){\stackrel{\ast}\rightarrow} (7,4)\stackrel{W}\rightarrow (6,4)$&(${\bf C}$)\\ 
$$&$(5,5)_{\Gamma, \Psi} $&:&$ (6,5)\stackrel{W}\rightarrow (5,5)$&(${\bf R}$)\\
$$&$(4,6)_{\Gamma} $&:&$ (4,7)\rightarrow (4,6)$&(${\bf H}$)\\ 
$$&$(4,6)_\Psi $&:&$ (5,8)\stackrel{\ast}\rightarrow (5,6)\stackrel{W}\rightarrow (4,6)$&(${\bf C}$)\\ 
$$&$(3,7)_{\Gamma, \Psi} $&:&$ (4,7)\stackrel{W}\rightarrow (3,7)$&(${\bf H}$)\\ 
$$&$(2,8)_{\Gamma} $&:&$ (2,9)\rightarrow (2,8)$&(${\bf R}$)\\
$$&$(2,8)_\Psi $&:&$ (5,8)\stackrel{\ast}\rightarrow (3,8)\stackrel{W}\rightarrow (2,8)$&(${\bf C}$)\\ 
$$&$(1,9)_{\Gamma ,\Psi}$&:&$(2,9)\stackrel{W}\rightarrow (1,9)$&(${\bf R}$)\\
$$&$(0,10)_{\Gamma} $&:&$ (0,11)\rightarrow (0,10)$&(${\bf H}$)\\ 
$$&$(0,10)_\Psi $&:&$(1,12)\stackrel{\ast}\rightarrow (1,10)\stackrel{W}\rightarrow (0,10)$&(${\bf C}$)\\ 
\hline
$D=11$&$(11,0)_{\Gamma}$&:&$ (13,0)\stackrel{\ast}\rightarrow (11,0)$&(${\bf C}$)\\ 
$$&$(11,0)_\Psi $&:&$ (13,0)\stackrel{W}\rightarrow (11,0)$&(${\bf H}$)\\
$$&$(10,1)_{\Gamma , \Psi} $&:&$ M.C.A.$&(${\bf R}$)\\ 
$$&$(9,2)_{\Gamma} $&:&$ (9,4)\stackrel{\ast}\rightarrow (9,2)$&(${\bf C}$)\\
$$&$(9,2)_\Psi $&:&$ (11,2)\stackrel{W}\rightarrow (9,2)$&(${\bf R}$)\\ 
$$&$(8,3)_{\Gamma ,\Psi}$&:&$ M.C.A.$&(${\bf H}$)\\
$$&$(7,4)_{\Gamma} $&:&$ (9,4)\stackrel{\ast}\rightarrow (7,4)$&(${\bf C}$)\\ 
$$&$(7,4)_\Psi $&:&$ (9,4)\stackrel{W}\rightarrow (7,4)$&(${\bf H}$)\\ 
$$&$(6,5)_{\Gamma, \Psi} $&:&$ M.C.A.$&(${\bf R}$)\\
$$&$(5,6)_{\Gamma} $&:&$ (5,8)\stackrel{\ast}\rightarrow (5,6)$&(${\bf C}$)\\ 
$$&$(5,6)_\Psi $&:&$ (7,6)\stackrel{W}\rightarrow (5,6)$&(${\bf R}$)\\ 
$$&$(4,7)_{\Gamma, \Psi} $&:&$ M.C.A.$&(${\bf H}$)\\ 
$$&$(3,8)_{\Gamma} $&:&$ (5,8)\stackrel{\ast}\rightarrow (3,8)$&(${\bf C}$)\\
$$&$(3,8)_\Psi $&:&$ (5,8)\stackrel{W}\rightarrow (3,8)$&(${\bf H}$)\\ 
$$&$(2,9)_{\Gamma ,\Psi}$&:&$M.C.A.$&(${\bf R}$)\\
$$&$(1,10)_{\Gamma} $&:&$ (1,12)\stackrel{\ast}\rightarrow (1,10)$&(${\bf C}$)\\ 
$$&$(1,10)_\Psi $&:&$ (3,10)\stackrel{W}\rightarrow (1,10)$&(${\bf R}$)\\ 
$$&$(0,11)_{\Gamma ,\Psi}$&:&$ M.C.A.$&(${\bf H}$)\\
\hline
\end{tabular}&\nonumber\\&&\end{eqnarray}}} }}

The three exceptional cases mentioned above, namely $(0,5)_{\Gamma}$, $(6,0)_{\Psi}$ and $(7,0)_{\Gamma}$, have been denoted by $\clubsuit$ in the table above.
They can be explicitly constructed as follows: the $(0,5)_\Gamma$ complex case can be recovered
from the quaternionic $(5,0)$ maximal Clifford algebra, which must be represented 
for this purpose through $4\times 4$ complex matrices, see \cite{crt1}, which we further multiply by the imaginary unit to
produce $(0,5)_\Gamma$. The knowledge of $(0,5)_\Gamma$ allows to produce,
by applying to it the second lifting algorithm of \cite{crt1},
the complex
$(7,0)_\Gamma$ irrep, given by $8\times 8$ complex matrices. At this point the
complex realization of  $(6,0)_\Psi$ fundamental spinors is obtained from the (Weyl-type) dimensional reduction $(7,0)_\Gamma\rightarrow (6,0)_\Psi$.\par
The construction specified by the table (\ref{maxclifford}) and supplemented by the three
exceptional cases listed above, allows us to reconstruct all Clifford algebra representations from the set of maximal Clifford algebra irreps. The explicit list of all the dimensional reductions of (non-maximal) Clifford irreps
and fundamental spinors for any $D$-dimensional space-time up to $D=11$ has been here produced. \par
The quaternionic Clifford irreps $(p,q)_\Gamma$ and the quaternionic fundamental spinor representations $(p,q)_\Psi$ can be realized through matrices with real, complex or quaternionic entries. On the other hand, the complex case (${\bf C}$) only allows Gamma matrices
representations in terms of matrices with either real or complex entries (in the real case
only matrices with real entries are allowed). In the following we will focus on generalized supersymmetries realized by either real or complex spinors (we leave the quaternionic 
spinors supersymmetries for future considerations). The complex supersymmetries can therefore be
constructed in association with both the ${\bf C}$ and ${\bf H}$ cases listed in table 
(\ref{gammapsi}).
For what concerns the real supersymmetries, they can be constructed without any restriction
for all the three (${\bf R}$, ${\bf C}$ and ${\bf H}$) cases. However, as pointed out in \cite{top}, in all the spacetimes supporting complex spinors, the corresponding real supersymmetries can be recovered from the most general complex supersymmetry. 
For this reason we found to be convenient to make a  
separate analysis of  the {\em purely} real supersymmetries (namely, the ones that cannot be recasted in
a complex spinor framework), which will be presented in Section {\bf 6}.

\section{Superparticles with tensorial central charges.}

In this section we review some of the basic ingredients entering the construction 
of the superparticle models with tensorial central charges. We present the most general action 
formulated with real spinors. Some subtleties and variants in the construction of the action
associated with complex spinors are elucidated and discussed in Section {\bf 11}.\par
Let us spend at first few words about the origin and the physical implications of the class of
models here presented. Rudychev and Sezgin \cite{rs} at first generalized the Brink-Schwarz
massless superparticles \cite{brsc} by introducing, for real spinors, a model associated with
the (\ref{Mgen}) superalgebra (their model, explicitly discussed in the $(2,2)$-spacetime,
can be considered as the simplest dynamical application of (\ref{Mgen})). Later Bandos and Lukierski
\cite{bl} reformulated the $(3+1)$ Minkowskian model (with $6$ additional bosonic tensorial
central charges) in terms of complex spinors and explicitly solved the constraints arising from
the Lagrange multipliers with the introduction of twistors. Their solution can be physically
interpreted \cite{bls} as a tower of massless higher spin particles, providing a dynamical framework for the Fronsdal's idea \cite{fro} about the role of the tensorial central charges (see the
discussion in \cite{sor} and references therein).\par
The most general action $S$ involving real spinors is constructed as follows \cite{rs} in
terms of the real superspace coordinates $X^{ab}$, $\Theta^a$ conjugated to
the superalgebra generators ${\cal Z}_{ab}$ and $Q_a$ of (\ref{Mgen}) ($X^{ab}$ is symmetric
in the $a\leftrightarrow b$ exchange). We have 
\begin{eqnarray}\label{realaction}
\relax S &=& \frac{1}{2}\int d\tau tr\left[{\cal Z}\cdot \Pi - e({\cal Z})^2\right],
\end{eqnarray}
where 
\begin{eqnarray}
\Pi^{ab} &=& dX^{ab} - \Theta^{(a}d\Theta^{b)},
\end{eqnarray}
while $e^{ab}$ denotes the Lagrange multipliers whose (anti)symmetry property is the
same as the one of the charge conjugation matrix $C^{ab}$, i.e.
\begin{eqnarray}
e^T=\varepsilon e \quad &for& \quad C^T=\varepsilon C.
\end{eqnarray}
By construction 
\begin{eqnarray}
{({\cal Z})^2}_{ab} &=& {\cal Z}_{ac} C^{cd} {\cal Z}_{db},
\end{eqnarray} 
namely the charge conjugation matrix is used as a metric to raise and lower spinorial indices.
\par
The massless constraint 
\begin{eqnarray}
({\cal Z})^2_{ab} &=& 0
\end{eqnarray}
is obtained from the variation $\delta e^{ab}$ of the Lagrange multipliers.\par
A symmetric charge conjugation matrix ($\varepsilon =1$) allows us \cite{rs} to construct a massive model
by simply performing a shift ${\cal Z} \rightarrow {\cal Z} + m C$ in the action 
(\ref{realaction}).\par
In order to introduce the action for the superparticle with complex spinors we mimick, as much as
possible, the real formulation. The bosonic matrix ${\cal Z}_{ab}$ is now replaced by
the pair of matrices ${\cal P}_{ab}$ and ${\cal R}_{a{\dot{b}}}$ (respectively symmetric and
hermitian) entering (\ref{Mhol}) and (\ref{Mher}). They can be accommodated in a symmetric matrix ${\bf P}$
(${\bf P}^T ={\bf P}$) as follows
\begin{eqnarray}\label{Pmatrix}
{\bf P} &=&\left(
\begin{tabular}{cc}
${\cal P}$ & ${\cal R}$\\
${\cal R}^\ast$ &${\cal P}^\ast$
\end{tabular}
\right).
\end{eqnarray}

The supercoordinates conjugated to ${\cal P}_{ab}$, ${\cal R}_{a{\dot b}}$, $Q_a$ and
${Q^\ast}_{\dot a}$ are given by $X^{ab}$, $Y^{a{\dot b}}$, $\Theta^a$ and ${{\Theta}^\ast}^{\dot a}$.\par
It is convenient to use the notation
\begin{eqnarray}
{\bf \Pi} &=&\left(
\begin{tabular}{cc}
$dX- \Theta d\Theta$ & $dY- \Theta d{\Theta}^\ast$\\
$dY^\ast - \Theta^\ast d\Theta$ &$dX^\ast -\Theta^\ast d\Theta^\ast$
\end{tabular}
\right).
\end{eqnarray}
We will also need the matrix
\begin{eqnarray}\label{Psquared}
{\bf P}^2 &=& {\bf P}{\cal C} {\bf P},
\end{eqnarray}
whose indices are raised by the metric ${\cal C}$. The available specific choices for ${\cal C}$ are discussed in Section {\bf 11}. The (anti)-symmetry property of ${\bf P}^2$ coincides
with the (anti)-symmetry property of ${\cal C}$.\par
The Lagrange multipliers enter a matrix
\begin{eqnarray}
{\bf E} &=&\left(
\begin{tabular}{cc}
$e$ & $f$\\
$g$ &$h$
\end{tabular}
\right).
\end{eqnarray}
In general, for any ${\bf U}$ (for our purposes
${\bf U}\equiv {\bf P}^2$) s.t. 
\begin{eqnarray}\label{Umatrix}
{\bf U} &=&\left(
\begin{tabular}{cc}
$U$ & $V$\\
$\lambda\mu V^\ast$ &$U^\ast$
\end{tabular}
\right)
\end{eqnarray}
with $U^T = \lambda U$, $V^\dagger = \mu V$ (therefore ${\bf U}^T=\lambda {\bf U}$),
the reality of the term
$tr ({\bf E U})$ requires
\begin{eqnarray}
g&=& \lambda\mu f^\ast,\nonumber\\
h&=& e^\ast.
\end{eqnarray}
A reality (imaginary) condition imposed on either ${\bf U}$ or ${\bf V}$ implies a
reality (imaginary) condition for the lagrange multipliers $e$ and $f$ respectively.\par
We are now in the position to write the action $S$ for the superparticle with bosonic tensorial
central charges and complex spinors as
\begin{eqnarray}\label{complexaction}
\relax S &=& \frac{1}{2}\int d\tau tr\left[{\bf P}{\bf \Pi} - {\bf E}({\bf P})^2\right].
\end{eqnarray}
As in the real case, a massive model can be introduced in correspondence of a symmetric ${\cal C}$ through the shift ${\bf P}\rightarrow {\bf P}+m{\cal C}$ in the action (\ref{complexaction}). 

\section{Generalized supersymmetries with real spinors.}

In this section we classify the generalized ``purely real" supersymmetries (the definition
of ``purely real" supersymmetries has been given at the end of Section {\bf 4}).
The complete list, up to $D=11$, of real Clifford irreps and real Clifford representations associated with fundamental spinors
is listed below
{\small
 { {{\begin{eqnarray}&\label{maxcliffreal}
\begin{tabular}{|l|l|}\hline

$D=1$&$(1,0), (0,1)_\Psi$\\ \hline
$D=2$&$(2,0)_\Gamma , (1,1)$\\ \hline
$D=3$&$(2,1), (1,2)_\Psi $\\ \hline
$D=4$&$(3,1)_\Gamma , (2,2) $\\ \hline
$D=5$&$ (3,2), (2,3)_\Psi   $\\ \hline
$D=6$&$ (4,2)_\Gamma , (3,3), (0,6)_\Gamma  $\\ \hline
$D=7$&$(7,0)_\Psi ,(4,3), (3,4)_\Psi , (0,7)   $\\ \hline
$D=8$&$ (8,0),(5,3)_\Gamma , (4,4), (1,7)_\Gamma , (0,8)   $\\ \hline
$D=9$&$ (9,0) ,(8,1)_\Psi , (5,4) , (4,5)_\Psi , (1,8), (0,9)_\Psi   $\\ \hline
$D=10$&$ (10,0)_\Gamma ,(9,1), (6,4)_\Gamma , (5,5), (2,8)_\Gamma , (1,9)   $\\ \hline
$D=11$&$ (10,1) ,(9,2)_\Psi , (6,5) , (5,6)_\Psi ,(2,9), (1,10)_\Psi   $\\ 

\hline

\end{tabular}&\nonumber\\&&\end{eqnarray}}} }} 

The above real Clifford representations can be obtained as reductions of the real maximal Clifford algebras satisfying $p-q = 1\quad mod\quad 8$, as specified by the following table
{\small
 { {{\begin{eqnarray}&\label{maxcliffreal2}
\begin{tabular}{|lll|}\hline

$(2,1)$&$\rightarrow$&$(0,1)_\Psi, (1,1), (2,0)_\Gamma $\\ \hline
$(3,2)$&$\rightarrow$&$(1,2)_\Psi , (2,2), (3,1)_\Gamma $\\ \hline
$(4,3)$&$\rightarrow$&$(2,3)_\Psi ,(3,3), (4,2)_\Gamma  $\\ 
$(0,7)$&$\rightarrow$&$ (0,6)_\Gamma $\\ \hline
$(5,4)$&$\rightarrow$&$ 
(3,4)_\Psi ,(4,4), (5,3)_\Gamma  $\\ 
$(1,8)$&$\rightarrow$&$(0,8) , (1,7)_\Gamma   $\\ 
$(9,0)$&$\rightarrow$&$ (7,0)_\Psi ,(8,0)   $\\ \hline
$(6,5)$&$\rightarrow$&$ (4,5)_\Psi , (5,5) , (6,4)_\Gamma   $\\ 
$(2,9)$&$\rightarrow$&$ (0,9)_\Psi ,(1,9), (2,8)_\Gamma    $\\
$(10,1)$&$\rightarrow$&$ (8,1)_\Psi ,(9,1) ,  (10,0)_\Gamma   $\\ \hline
$(11,2)$&$\rightarrow$&$ (9,2)_\Psi , (10,2) , (11,1)_\Gamma    $\\ 
$(7,6)$&$\rightarrow$&$ (5,6)_\Psi ,(6,6), (7,5)_\Gamma   $\\ 
$(3,10)$&$\rightarrow$&$ (1,10)_\Psi , (2,10) ,(3,9)_\Gamma   $\\ 

\hline

\end{tabular}&\nonumber\\&&\end{eqnarray}}} }} 

The (unique) charge conjugation matrix $C$ for maximal Clifford algebras
admits the following (anti)symmetry property, according to the dimension $D$ of
the space-time (its signature is not relevant)
{ {{\begin{eqnarray}&\label{realodd}
\begin{tabular}{|c|c|c|c|c|c|c|}\hline
$D=1$&$D=3$& $D=5$ &$D=7$&$D=9$&$D=11$&$D=13$\\ \hline
${\bf s}$&${\bf a}$& ${\bf a}$ &${\bf s}$&${\bf s}$&${\bf a}$&${\bf a}$
\\ \hline
\end{tabular}&\end{eqnarray}}} }

For a $D$-dimensional maximal Clifford algebra we can symbolically denote as $M_k$ the space
of $\left( \begin{array}{c}
  D\\
  k
\end{array}\right)$-component, totally antisymmetric, rank-$k$ tensors. The bosonic sector
of a ``saturated" generalized real supersymmetry (i.e. with maximal number, equal to $\frac{n(n+1)}{2}$, of bosonic components) for maximal Clifford algebras is given by
{ {{\begin{eqnarray}&\label{realodd2}
\begin{tabular}{|c|c|c|}\hline
spacetime&bosonic sectors&bosonic components\\ \hline
$D=1$&${M}_0$& $1$\\ \hline
$D=3$&${M}_1$& $3$\\ \hline
$D=5$&${{M}}_2$&$10$\\ \hline
$D=7$&${M}_0+{M}_3$&$1+35=36$\\ \hline
$D=9$&${M}_0+{M}_1+{M}_4$&$1+9+126=136$\\ \hline
$D=11$&${M}_1+{M}_2+{M}_5$&$11+55+462=528$\\ \hline
$D=13$&${M}_2+{M}_3+{M}_6$&$78+286+1716=2080$\\ \hline
\end{tabular}&\end{eqnarray}}} }
The dimensional reduction $D\rightarrow D-1$, corresponding to the signature passage $(p,q)\rightarrow (p,q-1)$ (here $D=p+q$), produces non-maximal Clifford algebras
whose decomposition into rank-$k$ antisymmetric tensors is symbolically denoted 
as ${\overline M}_k$ and is given as follows (we remind that this table corresponds
to a non-Weyl, $NW$-case)
{{\footnotesize{
{ {{\begin{eqnarray}&\label{realeven}
\begin{tabular}{|l|l|l|}\hline
spacetime&bosonic sectors&bosonic components\\ \hline
$D=3$&${M}_1\rightarrow \overline{M}_1+\overline{M}_0$& $3=2+1$\\ \hline
$D=5$&${{M}}_2\rightarrow \overline{M}_2+\overline{M}_1$&$10=6+4$\\ \hline
$D=7$&${M}_0+{M}_3\rightarrow \overline{M}_0+\overline{M}_3+
\overline{M}_2$&$36=1+20+15$\\ \hline
$D=9$&${M}_0+{M}_1+{M}_4\rightarrow 2\times\overline{M}_0+
\overline{M}_1+\overline{M}_4+
\overline{M}_3$&$136=2+8+70+56$\\ \hline
$D=11$&${M}_1+{M}_2+{M}_5\rightarrow \overline{M}_0+2\times\overline{M}_1+\overline{M}_2+
\overline{M}_4+\overline{M}_5$&$528=1+20+45+210+252$\\ \hline
$D=13$&${M}_2+{M}_3+{M}_6\rightarrow\overline{M}_1+2\times\overline{M}_2+\overline{M}_3
+\overline{M}_5+\overline{M}_6$&$2080=12+2\times 66+220+792+924$\\ \hline
\end{tabular}&\nonumber\\
&&\end{eqnarray}}} }  
}}
}
The dimensional reduction corresponding to the Weyl case $(p,q)\mapsto (p-1,q)$
is obtained by deleting a space-like Gamma matrix. Due to the presence of the Weyl projection,
only the bosonic components entering the upper-left block survive the dimensional reduction
(they are denoted in boldface in the table below). A factor $\frac{1}{2}$ is inserted
in order to remind that the corresponding rank-$\frac{D}{2}$ totally antisymmetric tensors
are self-dual. We have
{{
{ {{\begin{eqnarray}&\label{realevenweyl}
\begin{tabular}{|l|l|l|}\hline
spacetime&bosonic sectors&bosonic components\\ \hline
$D=2$&${M}_0+\frac{1}{2}{\bf{M}}_1$& $1$\\ \hline
$D=4$&$\frac{1}{2}{\bf{M}}_2+{M}_1$&$3$\\ \hline
$D=6$&${M}_0+\frac{1}{2}{\bf{M}}_3+{M}_2$&$10$\\ \hline
$D=8$&${\bf{M}}_0+{M}_1+{M}_3+\frac{1}{2}{\bf{M}}_4$&$36=1+35$\\ \hline
$D=10$&${M}_0+{\bf{M}}_1+{M}_2+{M}_4+\frac{1}{2}{\bf{M}}_5$&$136=10+126$\\ \hline
$D=12$&${M}_1+{\bf{M}}_2+{M}_3+{M}_5+2+\frac{1}{2}{\bf{M}}_6$&$528=66+462$\\ \hline
\end{tabular}&\nonumber\\
&&\end{eqnarray}}} }  
}}
The last dimensional reduction that we have to analyze is obtained by deleting
an extra space-like Gamma matrix, in order to produce the $(p-2,q)$ Weyl case,
whose decomposition in terms of rank-$k$ antisymmetric tensors is given by
{{
{ {{\begin{eqnarray}&\label{realevendoubleweyl}
\begin{tabular}{|l|llll|l|}\hline
$D=2   $&${{\widetilde M}_1}^{S.D.}$& $\rightarrow$&$D=1$&${\overline {M}}_0$&$1$\\ \hline
$D=4   $&${{\widetilde M}_2}^{S.D.}$& $\rightarrow$&$D=3$&${\overline {M}}_1$&                 $3$\\ \hline
$D=6   $&${{\widetilde M}_3}^{S.D.}$& $\rightarrow$&$D=5$&${\overline {M}}_2$&                   $10$\\ \hline
$D=8   $&${{\widetilde M}_0}+{{\widetilde M}_4}^{S.D.}$& $\rightarrow$&$D=7$&${\overline {M}}_0
+{\overline {M}}_3$&                      $36=1+35$\\ \hline
$D=10  $&${{\widetilde M}_1}+{{\widetilde M}_5}^{S.D.}$& $\rightarrow$&$D=9$&${\overline {M}}_0
+{\overline {M}}_1+{\overline {M}}_4$&                       $136=1+9+126$\\ \hline
$D=12  $&${{\widetilde M}_2}+{{\widetilde M}_6}^{S.D.}$& $\rightarrow$&$D=11$&$
{\overline {M}}_1+{\overline {M}}_2+{\overline {M}}_5$&                           $528=11+55+462$\\ \hline
\end{tabular}&\nonumber\\
&&\end{eqnarray}}} }  
}}
The above tables fully specify the whole set of saturated, purely real, generalized
supersymmetries up to $D=13$.

\section{Oxidation of real generalized supersymmetries and real superparticles with tensorial central charges.}

In this section we summarize some of the results obtained in the previous section concerning the
classification of saturated purely real generalized supersymmetries and we further apply them
to the construction and classification of the corresponding real superparticle models with
tensorial central charges.\par
It is convenient to list the (anti)symmetry properties of all types of matrices (the charge
conjugation matrices and those obtained, for non-maximal Clifford algebras, through the multiplication of external Gamma matrices) which can be used to define a metric for spinors.\par
We obtain, for non-maximal Clifford algebras arising from the given maximal Clifford algebras
entering the first column, the following table ($C$ denotes the charge conjugation matrix of
the associated maximal Clifford algebra)
{{
{ {{\begin{eqnarray}&\label{realcase1}
\begin{tabular}{|l|ll|ll|llll|}\hline
 $(p,q)$&$(p,q-1)$&&$(p-1,q)$&&$(p-2,q)$&&&\\ \hline \hline
 &$C$&$C\Gamma_t$&$C$&$C\Gamma_s$&$C$&$C\Gamma_{s_1}$&$C\Gamma_{s_2}$&$C\Gamma_{s_1}\Gamma_{s_2}$
 \\
 \hline \hline
  $(2,1)$&${\bf a}$&${\bf s}$&${\bf a}$&${\bf s}$&${\bf a}$&${\bf s}$&${{\bf s}}^W$&${{\bf s}}^W$\\ \hline
  
  $(3,2)$&${\bf a}$&${\bf a}$&${\bf a}^W$&${\bf a}^W$&${\bf a}^W$&${\bf a}^W$&${\bf a}$&${\bf a}$\\ \hline
  
  $(4,3)$&${\bf s}$&${\bf a}$&${\bf s}$&${\bf a}$&${\bf s}$&${\bf a}$&${\bf a}^W$&${\bf a}^W$\\ 
  
  $(0,7)$&${\bf s}$&${\bf a}$&$\times$&$\times$&$\times$&$\times$&$\times$&$\times$\\ \hline
  
  $(5,4)$&${\bf s}$&${\bf s}$&${\bf s}^W$&${\bf s}^W$&${\bf s}^W$&${\bf s}^W$&${\bf s}$&${\bf a}$\\ 
  
  $(1,8)$&${\bf s}$&${\bf s}$&${\bf s}^W$&${\bf s}^W$&$\times$&$\times$&$\times$&$\times$\\ 
  
  $(9,0)$&$\times$&$\times$&${\bf s}^W$&${\bf s}^W$&${\bf s}^W$&${\bf s}^W$&${\bf s}$&${\bf a}$\\ \hline
  
  $(6,5)$&${\bf a}$&${\bf s}$&${\bf a}$&${\bf s}$&${\bf a}$&${\bf s}$&${\bf s}^W$&${\bf s}^W$\\ 
  
  $(2,9)$&${\bf a}$&${\bf s}$&${\bf a}$&${\bf s}$&${\bf a}$&${\bf s}$&${\bf s}^W$&${\bf s}^W$\\ 
  
  $(10,1)$&${\bf a}$&${\bf s}$&${\bf a}$&${\bf s}$&${\bf a}$&${\bf s}$&${\bf s}^W$&${\bf s}^W$\\ \hline
  
  $(7,6)$&${\bf a}$&${\bf a}$&${\bf a}^W$&${\bf a}^W$&${\bf a}^W$&${\bf a}^W$&${\bf a}$&${\bf s}$\\ 
  
  $(3,10)$&${\bf a}$&${\bf a}$&${\bf a}^W$&${\bf a}^W$&${\bf a}^W$&${\bf a}^W$&${\bf a}$&${\bf s}$\\ 
 
  $(11,2)$&${\bf a}$&${\bf a}$&${\bf a}^W$&${\bf a}^W$&${\bf a}^W$&${\bf a}^W$&${\bf a}$&${\bf s}$\\ \hline
  
 \end{tabular}&\nonumber\\
&&\end{eqnarray}}} }  
}}
The $W$-suffix is introduced in a Weyl case in order to specify whether the corresponding 
matrix admits a non-vanishing upper-left block. If this is the case, it can be used as a metric
for Weyl-projected spinors of the same chirality (which has consequencies in the building of
the superparticle models).\par
Let us present now a table with the main properties of the real superparticle models associated with the purely real generalized supersymmetries. The following items are specified. The whole list (up to $D=13$)  of non-maximal purely real  Clifford irreps and fundamental spinors is presented in the second column (their associated maximal Clifford algebras are given in the first column). The third column, labeled by ``$\sharp$", specifies the number of components of the corresponding spinors, while ${\cal Z}$ and $e$ denote the number of independent bosonic components of, respectively, the matrix ${\cal Z}_{ab}$ and the lagrange multipliers entering the action (\ref{realaction}). The (anti)symmetry properties of the charge conjugation matrix $C$ associated with the maximal Clifford algebra are reported (${\widehat C}$ denotes the other choice for the charge conjugation matrix, for the spacetimes supporting two distinct charge conjugation matrices). ${\overline C}$ is the given charge conjugation matrix (if it exists) acting as metric for Weyl spinors of the same chirality (the symbol ``$\times$" is employed throughout the table if the corresponding case does not apply). As recalled in Section {\bf 5} the presence of a symmetric charge conjugation matrix allows to introduce a non-vanishing mass term in the superparticle model. The possibility ({\em yes-no}) of having a mass term is reported in the last column. We have 
{{
{ {{\begin{eqnarray}&\label{realcase2}
\begin{tabular}{|lllllllll|}\hline
 $MAX.$&$RED.$&$\sharp$&${\cal Z}$&$e$&$C$&${\widehat C}$&${\overline C}$&$mass$
 \\
 \hline
  $(2,1)$&$(2,0)_\Gamma$&$2$&$3$&$1$&${\bf a}$&$-$&$\times$&$no$\\ 
  
  $$&$(2,0)_\Gamma$&$2$&$3$&$3$&$-$&${\bf s}$&$\times$&$yes$\\ 
  
   $$&$(1,1)$&$1$&$\times$&$\times$&$\times$&$\times$&$\times$&$\times$\\
  
   $$&$(0,1)_\Psi$&$1$&$1$&$1$&$\times$&$\times$&${\bf s}$&$yes$\\ \hline
   
    $(3,2)$&$(3,1)_\Gamma$&$4$&$10$&$6$&${\bf a}$&$-$&$\times$&$no$\\ 
  
  $$&$(3,1)_\Gamma$&$4$&$10$&$6$&$-$&${\bf a}$&$\times$&$no$\\ 
  
   $$&$(2,2)$&$2$&$3$&$1$&$\times$&$\times$&${\bf a}$&$no$\\
  
   $$&$(1,2)_\Psi$&$2$&$3$&$1$&$\times$&$\times$&${\bf a}$&$no$\\ \hline
   
    $(4,3)$&$(4,2)_\Gamma$&$8$&$36$&$36$&${\bf s}$&$-$&$\times$&$yes$\\ 
  
  $$&$(4,2)_\Gamma$&$8$&$36$&$28$&$-$&${\bf a}$&$\times$&$no$\\ 
  
   $$&$(3,3)$&$4$&$\times$&$\times$&$\times$&$\times$&$\times$&$\times$\\
  
   $$&$(2,3)_\Psi$&$4$&$10$&$6$&$\times$&$\times$&${\bf a}$&$no$\\ \hline
   
    $(0,7)$&$(0,6)_\Gamma$&$8$&$36$&$36$&${\bf s}$&$-$&$\times$&$yes$\\ 
  
  $$&$(0,6)_\Gamma$&$8$&$36$&$28$&$-$&${\bf a}$&$\times$&$no$\\ 
   \hline
    $(5,4)$&$(5,3)_\Gamma$&$16$&$136$&$136$&${\bf s}$&$-$&$\times$&$yes$\\ 
  
  $$&$(5,3)_\Gamma$&$16$&$136$&$136$&$-$&${\bf s}$&$\times$&$yes$\\ 
  
   $$&$(4,4)$&$8$&$36$&$36$&$\times$&$\times$&${\bf s}$&$yes$\\
  
   $$&$(3,4)_\Psi$&$8$&$36$&$36$&$\times$&$\times$&${\bf s}$&$yes$\\ \hline

    $(6,5)$&$(6,4)_\Gamma$&$32$&$528$&$496$&${\bf a}$&$-$&$\times$&$no$\\ 
  
  $$&$(6,4)_\Gamma$&$32$&$528$&$528$&$-$&${\bf s}$&$\times$&$yes$\\ 
  
   $$&$(5,5)$&$16$&$\times$&$\times$&$\times$&$\times$&$\times$&$\times$\\
  
   $$&$(4,5)_\Psi$&$16$&$136$&$136$&$\times$&$\times$&${\bf s}$&$yes$\\ \hline
   
   $(7,6)$&$(7,5)_\Gamma$&$64$&$2080$&$2016$&${\bf a}$&$-$&$\times$&$no$\\ 
  
  $$&$(7,5)_\Gamma$&$64$&$2080$&$2016$&$-$&${\bf a}$&$\times$&$no$\\ 
  
   $$&$(6,6)$&$32$&$528$&$496$&$\times$&$\times$&${\bf a}$&$no$\\
  
   $$&$(5,6)_\Psi$&$32$&$528$&$496$&$\times$&$\times$&${\bf a}$&$no$\\ \hline

 \end{tabular}&\nonumber\\
&&\end{eqnarray}}} }  
}}
In the above table only the minimal constructions of real generalized supersymmetries
are classified. Non-minimal cases can be obtained by using more than one copy of
spinors. The Weyl cases $(1,1)$, $(3,3)$ and $(5,5)$, e.g., are not present because
the corresponding charge conjugation matrices have a vanishing upper left block,
see the table ({\ref{realcase1}). As a consequence, in these spacetimes, Weyl spinors of both chiralities have to be introduced in the analog of action (\ref{realaction}), making the whole construction non-minimal, in the sense specified above.\par
The table ({\ref{realcase2}) summarizes the main properties of purely real superparticle models with tensorial
central charges. In the next sections we will discuss the construction of the superparticles
obtained from complex spinors. It involves many subtleties not present in the real construction. 

\section{Generalized supersymmetries with complex spinors.}

Let us now introduce the generalized supersymmetries constructed with complex spinors,
whose basic algebraic relations are given by ({\ref{Mhol}) and (\ref{Mher}).\par
Clifford representations supporting complex spinors are obtained by the quaternionic
(using the fact that  quaternions can be represented as complex matrices) maximal
Clifford algebras satisfying the $p-q=5\quad mod\quad 8$ condition (see the discussion in Section {\bf 4}). The complete list of reductions leading to the non-maximal complex
representations of type $(p,q)_\Gamma$ and $(p,q)_\Psi$ is furnished in the next Section.\par
We present here the decomposition of the bosonic sector (in terms of the rank-$k$ totally 
antisymmetric tensors denoted as $M_k$) for the $D$-dimensional (up to $D=13$) complex generalized supersymmetries in maximal Clifford algebras spacetimes. We get the following table,
where the complex size ($n_{\bf C}$) of the Clifford Gamma matrices is reported in the second column, while the number of bosonic components and the bosonic sectors are split into two
parts, entering respectively the matrices ${\cal P}$ and ${\cal R}$ of formulas 
(\ref{Mhol}) and (\ref{Mher}).
{\small
{ {{\begin{eqnarray}&\label{complexmaximal}
\begin{tabular}{|c|c||c|c||c|c||}\hline
spacetime&$n_{\bf C}$& sym. b. c.& sym. bos. sect.&her. b.c.& her. bos. sect.\\ \hline
$D=3$&$2$&$6$&$2{M}_1$&$4$& $M_0+M_1$\\ \hline
$D=5$&$4$&$20$&$2{M}_2$& $16$&$M_0+M_1+M_2$\\ \hline
$D=7$&$8$&$72$&$2({{M}}_0+M_3)$&$64$&$M_0+\ldots +M_3$\\ \hline
$D=9$&$16$&$272$&$2({M}_0+{M}_1+M_4)$&$256$&$M_0+\ldots +M_4$\\ \hline
$D=11$&$32$&$1056$&$2({M}_1+{M}_2+{M}_5)$&$1024$&$M_0+\ldots +M_5$\\ \hline
$D=13$&$64$&$4160$&$2({M}_2+{M}_3+{M}_6)$&$4096$&$M_0+\ldots +M_6$\\ \hline
\end{tabular}&\nonumber \\\end{eqnarray}}} }
}
The above table is easily computed with the knowledge, according to the discussion at the beginning of Section {\bf 3}, of the symmetric (${\bf s}$) or antisymmetric (${\bf a}$) property
of the unique charge conjugation matrix $C$, as well as the hermitian (${\bf +}$) or antihermitian (${\bf -}$) property of the $A$ matrix. We have, for the above $D$-dimensional spacetimes
associated with maximal Clifford algebras,
{ {{\begin{eqnarray}&\label{canda}
\begin{tabular}{|c|c|c|c|c|c|}\hline
$D=3$& $D=5$ &$D=7$&$D=9$&$D=11$&$D=13$\\ \hline
\begin{tabular}{cc}
$C$&$A$\\
${\bf a}$&${\bf +}$
\end{tabular}
& \begin{tabular}{cc}
$C$&$A$\\
${\bf a}$&${\bf +}$
\end{tabular}
& \begin{tabular}{cc}
$C$&$A$\\
${\bf s}$&${\bf -}$
\end{tabular}
& \begin{tabular}{cc}
$C$&$A$\\
${\bf s}$&${\bf -}$
\end{tabular}
& \begin{tabular}{cc}
$C$&$A$\\
${\bf a}$&${\bf +}$
\end{tabular}
& \begin{tabular}{cc}
$C$&$A$\\
${\bf a}$&${\bf +}$
\end{tabular}
\\ \hline
\end{tabular}&\end{eqnarray}}} }
It is worth noticing that the most general, ``saturated", complex generalized supersymmetry
is obtained by adding the symmetric and hermitian bosonic sectors of the table (\ref{complexmaximal}).
As an example, in the $D=5$ case, the full bosonic sector is given by the
$1+5+3\times 10=36$ components entering 
\begin{eqnarray}
\label{example}
&M_0+M_1+3 M_2&
\end{eqnarray} (and similarly for the other cases). It is worth noticing that the saturated cases can be reproduced also in terms of real spinors \cite{top}. The way of counting is different, but at the end the same results are reproduced.
As an example, the $D=5$ case for real spinors can be obtained as
dimensional reduction of the $D=7$ real maximal Clifford algebra. The $D=7$ dimensional
$1+35=36$  bosonic sector $M_0+M_3$ gets decomposed, in the $D=5$-dimensional viewpoint, precisely as (\ref{example}). \par
The introduction of a formulation, whenever this is possible, based on complex spinors,
is however of fundamental importance for what concerns the construction of the constrained generalized supersymmetries discussed in Section 
{\bf 10}. The constrained supersymmetries cannot be recovered within the real framework.
They can only be understood within the setting based on complex spinors. 

\section{Oxidation of complex generalized supersymmetries and tables of results.}

In order to complete the classification of the complex generalized supersymmetries
we have to specify how to recover the non-maximal cases in terms of their oxidized form.
The results of this section parallel, for the complex case, those of Section {\bf 6}, which
was devoted to the real case. In the main tables below we list the following properties, starting from the given maximal Clifford algebras which admit complex spinors.
We present the full list of non-maximal, reduced $(p,q)_\Gamma$ and $(p,q)_\Psi$ cases.
The (anti)symmetry properties (${\bf s}, {\bf a}$) and the (anti)hermitian properties
(${\bf +}, {\bf -}$ as in the previous section) for the matrices $C$ and $A$ of the maximal
Clifford algebras and those obtained by multiplication of the external Gamma matrices (in the reduced case) are given.
It is worth reminding, see the discussion in Section {\bf 4}, that for the four complex cases 
$(p,q-2)_\Gamma$, $(p-2,q)_\Gamma$, $(p-1,q-2)_\Psi$, $(p-3,q)_\Psi$ the two deleted Gamma
matrices associated with the imaginary quaternions do not appear in the tables below.\par
As in table (\ref{realcase1}) we put a suffix $W$ to denote whether, in the Weyl case, the corresponding matrices admit a non-vanishing upper-left block (as before, they can be used as a metric for the Weyl-projected spinors of the same chirality).
The series of reductions in the non-Weyl ($NW$)-case are given by
{{
{ {{\begin{eqnarray}&
\begin{tabular}{|l|llll|ll|ll|}\hline \label{compcase1}
 $(p,q)$&&$(p,q-1)_\Gamma$&&&$(p,q-2)_\Gamma$&&$(p-2,q)_\Gamma$&\\ \hline \hline
 & $C$&$C\Gamma_t$&$A$&$A\Gamma_t$&$C$&$A$&$C$&$A$
 \\
 \hline \hline
  $(0,3)$&${\bf a}$&${\bf s}$&${\bf +}$&${\bf -}$&${\bf s}$&${\bf -}$&$\ldots$&$\ldots$\\ \hline
  
  $(1,4)$&${\bf a}$&${\bf a}$&${\bf +}$&${\bf +}$&${\bf a}$&${\bf -}$&$\ldots$&$\ldots$\\ 
  
  $(5,0)$&${\bf a}$&$\ldots$&${\bf +}$&$\ldots$&$\ldots $&$\ldots$&${\bf a}$&${\bf +}$\\ \hline
  
  $(2,5)$&${\bf s}$&${\bf a}$&${\bf -}$&${\bf +}$&${\bf a}$&${\bf +}$&$\ldots$&$\ldots$\\ 
  
  $(6,1)$&${\bf s}$&${\bf a}$&${\bf -}$&${\bf +}$&$\ldots$&$\ldots$&${\bf a}$&${\bf -}$\\ \hline
  
  $(3,6)$&${\bf s}$&${\bf s}$&${\bf -}$&${\bf -}$&${\bf s}$&${\bf +}$&${\bf s}$&${\bf -}$\\ 
  
  $(7,2)$&${\bf s}$&${\bf s}$&${\bf -}$&${\bf -}$&${\bf s}$&${\bf +}$&${\bf s}$&${\bf -}$\\ \hline
  
  $(4,7)$&${\bf a}$&${\bf s}$&${\bf +}$&${\bf -}$&${\bf s}$&${\bf -}$&${\bf s}$&${\bf +}$\\ 
  
  $(8,3)$&${\bf a}$&${\bf s}$&${\bf +}$&${\bf -}$&${\bf s}$&${\bf -}$&${\bf s}$&${\bf +}$\\ 
  
  $(0,11)$&${\bf a}$&${\bf s}$&${\bf +}$&${\bf -}$&${\bf s}$&${\bf -}$&$\ldots$&$\ldots$\\ \hline
  
  $(5,8)$&${\bf a}$&${\bf a}$&${\bf +}$&${\bf +}$&${\bf a}$&${\bf -}$&${\bf a}$&${\bf +}$\\ 
  
  $(9,4)$&${\bf a}$&${\bf a}$&${\bf +}$&${\bf +}$&${\bf a}$&${\bf -}$&${\bf a}$&${\bf +}$\\ 
 
  $(1,12)$&${\bf a}$&${\bf a}$&${\bf +}$&${\bf +}$&${\bf a}$&${\bf -}$&$\ldots$&$\ldots$\\ 
  
  $(0,13)$&${\bf a}$&${\bf a}$&${\bf +}$&${\bf +}$&${\bf a}$&${\bf -}$&$\ldots$&$\ldots$\\ \hline
  
 \end{tabular}&\\ \nonumber
&&\end{eqnarray}}} }  
The reductions associated to the Weyl case ($W$-case) are expressed by the two following tables. We have
}}
{\footnotesize
{{
{ {{\begin{eqnarray}&\label{compcase2}
\begin{tabular}{|l|llll|llllllll|}\hline
 $(p,q)$&&$(p-1,q)$&&&&&&$(p-2,q)_\Psi$&&&&\\ \hline \hline
 &$C$&$C\Gamma_s$&$A$&$A\Gamma_s$&$C$&$C\Gamma_{s1}$&$C\Gamma_{s2}$&
 $C\Gamma_{s1}\Gamma_{s2}$&$A$&$A\Gamma_{s1}$&$A\Gamma_{s2}$&
 $A\Gamma_{s1}\Gamma_{s2}$
 \\
 \hline \hline
 
  $(1,4)$&${\bf a}^W$&${\bf a}^W$&${\bf +}^W$&${\bf +}^W$&$\ldots$&$\ldots$&$\ldots$&$\ldots$&$\ldots$&$\ldots$&$\ldots$&$\ldots$\\ 
  
  $(5,0)$&${\bf a}^W$&${\bf a}^W$&${\bf +}^W$&${\bf +}^W$
  &${\bf a}^W$&${\bf a}^W$&${\bf a}$&${\bf s}$
  &${\bf +}^W$&${\bf +}^W$&${\bf +}$&${\bf -}$\\ \hline
  
  $(2,5)$&${\bf s}$&${\bf a}$&${\bf -}$&${\bf +}$
  &${\bf s}$&${\bf a}$&${\bf a}^W$&${\bf a}^W$
  &${\bf -}$&${\bf +}$&${\bf +}^W$&${\bf +}^W$\\ 
  
  $(6,1)$&${\bf s}$&${\bf a}$&${\bf -}$&${\bf +}$
  &${\bf s}$&${\bf a}$&${\bf a}^W$&${\bf a}^W$
  &${\bf -}$&${\bf +}$&${\bf +}^W$&${\bf +}^W$
  \\ \hline
  
  $(3,6)$&${\bf s}^W$&${\bf s}^W$&${\bf -}^W$&${\bf -}^W$
  &${\bf s}^W$&${\bf s}^W$&${\bf s}$&${\bf a}$
  &${\bf -}^W$&${\bf -}^W$&${\bf -}$&${\bf +}$

\\ 
  
  $(7,2)$&${\bf s}^W$&${\bf s}^W$&${\bf -}^W$&${\bf -}^W$
  &${\bf s}^W$&${\bf s}^W$&${\bf s}$&${\bf a}$
  &${\bf -}^W$&${\bf -}^W$&${\bf -}$&${\bf +}$
  \\ \hline
  
  $(4,7)$&${\bf a}$&${\bf s}$&${\bf +}$&${\bf -}$
  &${\bf a}$&${\bf s}$&${\bf s}^W$&${\bf s}^W$
  &${\bf +}$&${\bf -}$&${\bf -}^W$&${\bf -}^W$\\ 
  
  $(8,3)$&${\bf a}$&${\bf s}$&${\bf +}$&${\bf -}$
  &${\bf a}$&${\bf s}$&${\bf s}^W$&${\bf s}^W$
  &${\bf +}$&${\bf -}$&${\bf -}^W$&${\bf -}^W$
  
  \\ 
   \hline
  
  $(5,8)$&${\bf a}^W$&${\bf a}^W$&${\bf +}^W$&${\bf +}^W$
  &${\bf a}^W$&${\bf a}^W$&${\bf a}$&${\bf s}$
  &${\bf +}^W$&${\bf +}^W$&${\bf +}$&${\bf -}$\\ 
  
  $(9,4)$&${\bf a}^W$&${\bf a}^W$&${\bf +}^W$&${\bf +}^W$
  &${\bf a}^W$&${\bf a}^W$&${\bf a}$&${\bf s}$
  &${\bf +}^W$&${\bf +}^W$&${\bf +}$&${\bf -}$\\ 
 
  $(1,12)$&${\bf a}^W$&${\bf a}^W$&${\bf +}^W$&${\bf +}^W$
  &$\ldots$&$\ldots$&$\ldots$&$\ldots$
  &$\ldots$&$\ldots$&$\ldots$&$\ldots$\\ 
  
  \hline
  
 \end{tabular}&\nonumber\\
&&\end{eqnarray}}} }  
}}
}
and
{\footnotesize{
{ {{\begin{eqnarray}&\label{compcase3}
\begin{tabular}{|l|llll|llll|}\hline
 $(p,q)$&&$(p-1,q-2)_\Psi$&&&&$(p-3,q)_\Psi$&&\\ \hline \hline
 & $C$&$C\Gamma_s$&$A$&$A\Gamma_s$&$C$&$C\Gamma_s$&$A$&$A\Gamma_s$
 \\
 \hline \hline
 
  $(1,4)$&${\bf a}$&${\bf s}$&${\bf -}^W$&${\bf -}^W$&$\ldots$&$\ldots$&$\ldots$&$\ldots$\\ 
  
  $(5,0)$&$\ldots$&$\ldots$&$\ldots$&$\ldots$&${\bf a}$&${\bf s}$&${\bf +}^W$&${\bf +}^W$\\ \hline
  
  $(2,5)$&${\bf a}^W$&${\bf a}^W$&${\bf +}$&${\bf -}$&$\ldots$&$\ldots$&$\ldots$&$\ldots$\\ 
  
  $(6,1)$&$\ldots$&$\ldots$&$\ldots$&$\ldots$&${\bf a}^W$&${\bf a}^W$&${\bf -}$&${\bf +}$\\ \hline
  
  $(3,6)$&${\bf s}$&${\bf a}$&${\bf +}^W$&${\bf +}^W$&${\bf s}$&${\bf a}$&${\bf -}^W$&${\bf -}^W$\\ 
  
  $(7,2)$&${\bf s}$&${\bf a}$&${\bf +}^W$&${\bf +}^W$&${\bf s}$&${\bf a}$&${\bf -}^W$&${\bf -}^W$\\ \hline
  
  $(4,7)$&${\bf s}^W$&${\bf s}^W$&${\bf -}$&${\bf +}$&${\bf s}^W$&${\bf s}^W$&${\bf +}$&${\bf -}$\\ 
  
  $(8,3)$&${\bf s}^W$&${\bf s}^W$&${\bf -}$&${\bf +}$&${\bf s}^W$&${\bf s}^W$&${\bf +}$&${\bf -}$\\ 
   \hline
  
  $(5,8)$&${\bf a}$&${\bf s}$&${\bf -}^W$&${\bf -}^W$&${\bf a}$&${\bf s}$&${\bf +}^W$&${\bf +}^W$\\ 
  
  $(9,4)$&${\bf a}$&${\bf s}$&${\bf -}^W$&${\bf -}^W$&${\bf a}$&${\bf s}$&${\bf +}^W$&${\bf +}^W$\\ 
 
  $(1,12)$&${\bf a}$&${\bf s}$&${\bf -}^W$&${\bf -}^W$&$\ldots$&$\ldots$&$\ldots$&$\ldots$\\ 
  
  \hline
  
 \end{tabular}&\nonumber\\
&&\end{eqnarray}}} }  
}}
It is convenient to extract from the previous tables and separately present the list of types
of metric ((anti)symmetric and/or (anti)hermitian) which can be introduced on complex spinors
in correspondence with each one of the complex structure supporting the $(p,q)_{\Gamma,\Psi}$ representations of the Clifford algebras. We have
{{
{ {{\begin{eqnarray}&\label{compcase4}
\begin{tabular}{|ll|ll|ll|}\hline
 $(3,0)_\Gamma:$&$({\bf a+})$&$(6,1):$&$({\bf s-})$&$(10,0)_\Psi:$
 &$({\bf +})^W$\\
 
 $(3,0)_\Psi:$&$({\bf a+})^W$&$(5,2)_\Gamma:$&$({\bf s-})$&$(9,2)_\Gamma:$
 &$({\bf sa,+-})$\\
 
 $(1,2)_\Gamma:$&$({\bf a-})$&$(5,2)_\Psi:$&$({\bf s-})^W$&$(8,2)_\Psi:$
 &$({\bf -})^W$\\
 
 $(0,3):$&$({\bf a+})$&$(3,4)_\Gamma:$&$({\bf s+})$&$(7,3):$
 &$({\bf sa,+-})$\\
 
 $(4,0):$&$({\bf a+})^W$&$(2,5):$&$({\bf s-})$&$(6,4)_\Psi:$
 &$({\bf +})^W$\\
 
 $(3,1)_\Psi:$&$({\bf a})^W$&$(1,6)_\Gamma:$&$({\bf s-})$&$(4,6)_\Gamma:$
 &$({\bf sa,+-})$\\
 
 $(1,3)_\Gamma:$&$({\bf a+})$&$(1,6)_\Psi:$&$({\bf s-})^W$&$(4,6)_\Psi:$
 &$({\bf -})^W$\\
 
 $(1,3)_\Psi:$&$({\bf a})^W$&$(7,1)_\Gamma:$&$({\bf s-})$&$(3,7):$
 &$({\bf sa,+-})$\\
 
 $(0,4):$&$({\bf a+})^W$&$(7,1)_\Psi:$&$({\bf s})^W$&$(2,8)_\Psi:$
 &$({\bf +})^W$\\
 
 $(5,0):$&$({\bf a+})$&$(6,2):$&$({\bf s-})^W$&$(0,10)_\Gamma:$
 &$({\bf sa,+-})$\\
 
 $(4,1)_\Gamma:$&$({\bf a-})$&$(5,3)_\Psi:$&$({\bf s})^W$&$(0,10)_\Psi:$
 &$({\bf -})^W$\\
 
 $(4,1)_\Psi:$&$({\bf a+})^W$&$(3,5)_\Gamma:$&$({\bf s-})$&$(11,0)_\Gamma:$
 &$({\bf a-})$\\
 
 $(2,3)_\Gamma:$&$({\bf a+})$&$(3,5)_\Psi:$&$({\bf s})^W$&$(11,0)_\Psi:$
 &$({\bf a+})^W$\\
 
 $(1,4):$&$({\bf a+})$&$(2,6):$&$({\bf s-})^W$&$(9,2)_\Gamma:$
 &$({\bf a-})$\\
 
 $(0,5)_\Gamma:$&$({\bf a-})$&$(1,7)_\Psi:$&$({\bf s})^W$&$(8,3):$
 &$({\bf a+})$\\
 
 $(0,5)_\Psi:$&$({\bf a+})^W$&$(8,1)_\Gamma:$&$({\bf s-})$&$(7,4)_\Gamma:$
 &$({\bf a+})$\\
 
 $(6,0)_\Gamma:$&$({\bf sa,+-})$&$(7,2):$&$({\bf s-})$&$(7,4)_\Psi:$
 &$({\bf a+})^W$\\
 
 $(6,0)_\Psi:$&$({\bf +})^W$&$(6,3)_\Gamma:$&$({\bf s+})$&$(5,6)_\Gamma:$
 &$({\bf a-})$\\
 
 $(5,1):$&$({\bf sa,+-})$&$(6,3)_\Psi:$&$({\bf s-})^W$&$(4,7):$
 &$({\bf a+})$\\
 
 $(4,2)_\Psi:$&$({\bf -})^W$&$(4,5)_\Gamma:$&$({\bf s-})$&$(3,8)_\Gamma:$
 &$({\bf a+})$\\
 
 $(2,4)_\Gamma:$&$({\bf sa,+-})$&$(3,6)_\Gamma:$&$({\bf s-})$&$(3,8)_\Psi:$
 &$({\bf a+})^W$\\
 
 $(2,4)_\Psi:$&$({\bf +})^W$&$(3,6)_\Psi:$&$({\bf s})^W$&$(1,10)_\Gamma:$
 &$({\bf a-})$\\
 
 $(1,5):$&$({\bf sa,+-})$&$(2,7)_\Gamma:$&$({\bf s+})$&$(0,11):$
 &$({\bf a-})$\\
 
 $(0,6)_\Psi:$&$({\bf -})^W$&$(0,7)_\Psi:$&$({\bf s-})^W$&$$
 &$$\\
 
 $(7,0)_\Gamma:$&$({\bf sa,+-})$&$(0,9)_\Gamma:$&$({\bf s-})$&$$
 &$$\\
 
  \hline
 \end{tabular}&\nonumber\\
&&\end{eqnarray}}} }  
}}
Some remarks on the previous table are in order. It should be noticed that some
given Clifford representations admit different metrics of opposite symmetry or opposite
hermiticity. The $W$-suffix has been introduced, as above, to denote whether in the Weyl case
the corresponding metric is non-vanishing in the upper-left block and can be used as a metric
for Weyl spinors of the same chirality (i.e. for the minimal theories, see the discussion in
the previous section).\par
The information contained in (\ref{compcase4}) allows to reconstruct the properties of the corresponding 
generalized supersymmetries, in full analogy with the real case. In the following we will
discuss an example of application of the results produced in this section in the context of the superparticles with tensorial central charges.

\section{Constrained generalized supersymmetries and their duality relations.}

In this section we investigate and classify the set of consistent constraints that can be imposed on the complex generalized supersymmetries.\par
We already recalled at the end of Section {\bf 8} the results of a discussion in \cite{top}, namely
that the saturated complex generalized supersymmetries (i.e. the ones admitting as bosonic r.h.s. both the most general symmetric matrix ${\cal P}$ entering (\ref{Mhol}) and the most general hermitian matrix
${\cal R}$ entering (\ref{Mher})) contain the same number of bosonic degrees of freedom as the corresponding saturated generalized supersymmetries realized with {\em real} spinors. In this
respect the big advantage of the introduction of the complex formalism, whenever this is indeed possible, consists in the implementation of some constraint that cannot be otherwise imposed within the real framework.\par
In \cite{top} the two big classes of hermitian and holomorphic generalized supersymmetries were introduced and discussed. In this work we extend such a result by presenting a whole new class of division-algebra related constraints that can be consistently imposed. The bosonic r.h.s.
can be expressed in terms of the rank-$k$ totally antisymmetric tensors (denoted as $M_k$),
see Section  {\bf 3}. It is clear that any restriction on the saturated bosonic generators which allows all
possible combinations of the rank-$k$ antisymmetric tensors entering the r.h.s.  is in principle admissible by a Lorentz-covariant requirement\footnote{In this work we are limiting our discussion on the generalized supersymmetries which can be loosely denoted as ``generalized supertranslations", see \cite{top}. Supersymmetries of this kind present no Lorentz generators. However, they can be regarded as building blocks to construct superconformal algebras, out of
which the generalized superPoincar\'e algebras, admitting Lorentz subalgebras, can be recovered
through an Inon\"{u}-Wigner type of contraction. The procedure to construct the associated 
superconformal algebra starting from a given ``generalized supertranslation algebra" has been
illustrated in \cite{lt1} and further discussed in \cite{top}. It requires the introduction of two separated
copies of ``generalized supertranslations". The implementation of super-Jacobi identities is
sufficient to detect the remaining generators and close the whole set of algebraic
relations defining the associated superconformal algebra. Therefore, all the information about such superconformal algebras is already contained in the generalized supertranslations, the subject of the present investigation and classification.}.
On the other hand few particular combinations of the rank-$k$ antisymmetric tensors have more compelling reasons to appear than just arising as a hand-imposed restriction on the saturated bosonic r.h.s. They can indeed be present due to a division-algebra constraint based on an underlying
symmetry. It is expected that restrictions of this type offer a protecting mechanism towards the arising of anomalous terms, in application to the supersymmetries realized by certain classes of dynamical systems. This is an important reason to analyze and classify these constraints. Their whole class is presented in the table below. It consists of all possible combinations of restrictions on the ${\cal P}$, ${\cal R}$ matrices of (\ref{Mhol}) and (\ref{Mher}) (e.g. whether both of them are present or just one of them, if a reality or an imaginary condition is applied).
The entries in the table below specify the number of bosonic components (in the real counting)
associated with the given constrained supersymmetry realized by $n$-component complex spinors.
The columns represent the restrictions on ${\cal R}$, the rows the restrictions on ${\cal P}$
(an imaginary condition on ${\cal P}$ is equivalent to the reality condition and therefore is not reported in the table below). We have   
{{
{ {{\begin{eqnarray}&
\begin{tabular}{|c|c|c|c|c|}\hline \label{constrained}
 ${\cal P}\backslash {\cal R} $&$1)\quad Full$&$2)\quad Real$&$3)\quad Imag.$&$4)\quad Abs.$
 \\
 \hline
  $a)\quad Full$&$2n^2+n$&$\frac{3}{2}(n^2+n)$&$\frac{1}{2}(3n^2+n)$&$n^2+n$\\ \hline
  
  $b)\quad Real$&$\frac{1}{2}(3n^2+n)$&$n^2+n$&$n^2$&$\frac{1}{2}(n^2+n)$\\ \hline
  
   $c)\quad Abs.$&$n^2$&$\frac{1}{2}(n^2+n)$&$\frac{1}{2}(n^2-n)$&$0$\\
  
  \hline

 \end{tabular}&\\ \nonumber
&&\end{eqnarray}}} }  
}}
Some comments are in order. The above list of constraints is not necessarily implemented 
for any given supersymmetric dynamical system. One should check, e.g., that the above
restrictions are indeed compatible with the equations of motion. On a purely algebraic
basis, however, they are admissible restrictions which require a careful investigation.\par
One can notice that certain numbers appear twice as entries in the above table.
This is related with the fact that the same constrained superalgebra can admit a different,
but equivalent, presentation. We refer to these equivalent presentations as ``dual formulations" of the constrained supersymmetries. Dual formulations are expected
in correspondence of the constraints
\begin{eqnarray}
a3 &\leftrightarrow& b1,\nonumber\\
a4 &\leftrightarrow& b2,\nonumber\\
b3 &\leftrightarrow& c1,\nonumber\\
b4 &\leftrightarrow& c2.
\end{eqnarray}
It is worth stressing that in application to dynamical systems, which need more data
than just superalgebraic data, one should explicitly verify whether the above related constraints indeed lead to equivalent theories.\par
The inequivalent constrained generalized supersymmetries can be listed as follows
{{
{ {{\begin{eqnarray}&
\begin{tabular}{|c|clll|}\hline \label{dualities}
 
  $I$&$(a1)$&$2n^2+n,$&$k=3,$&$l=1$\\\hline
  $II$&$(a2)$&$\frac{3}{2}(n^2+n),$&$k=3,$&$l=0$\\ \hline
  $III$&$(a3\, \&\, b1)$&$\frac{1}{2}(3n^2+n),$&$k=2,$&$l=1$\\ \hline
  $IV$& $(a4\, \&\, b2)$&$n^2+n,$&$k=2,$&$l=0$\\ \hline
  $V$&$(b3\, \& \, c1)$&$n^2,$&$k=1,$&$l=1$\\ \hline
  $VI$&$(b4\, \&\, c2)$&$\frac{1}{2}(n^2+n),$&$k=1,$&$l=0$\\ \hline
  $VII$&$(c3)$&$\frac{1}{2}(n^2-n),$&$k=0,$&$l=1$\\ \hline

 \end{tabular}&\\ \nonumber
&&\end{eqnarray}}} }  
}}
The integral numbers $k,l$ have the following meaning. For the given constrained
supersymmetry the bosonic r.h.s.
can be presented in the following form 
\begin{eqnarray}\label{constrainedsectors}
Z&=& k X + lY, \quad k=0,1,2,3,\quad l=0,1,
\end{eqnarray}
where $X$ and $Y$ denote the bosonic sectors associated with the
$VI$ and respectively $VII$ constrained supersymmetry.\par
In association with the maximal Clifford algebras in $D$-dimensional spacetimes
(with no dependence on their signature), the $X$ and $Y$ bosonic sectors are given
by the following set of rank-$k$ antisymmetric tensors
{{
{ {{\begin{eqnarray}&
\begin{tabular}{|c|c|c|}\hline \label{dualities2}
  &$X$&$Y$\\ \hline
  $D=3$&$M_1$&$M_0$\\ \hline
    $D=5$&$M_2$&$M_0+M_1$\\ \hline
      $D=7$&$M_0+M_3$&$M_1+M_2$\\ \hline
        $D=9$&$M_0+M_1+M_4$&$M_2+M_3$\\ \hline
          $D=11$&$M_1+M_2+M_5$&$M_0+M_3+M_4$\\ \hline
            $D=13$&$M_2+M_3+M_6$&$M_0+M_1+M_4+M_5$\\ \hline  
 \end{tabular}&\\ \nonumber
&&\end{eqnarray}}} }  
}}
Formula (\ref{constrainedsectors}) specifies the admissible class of division-algebra related, constrained
bosonic sectors.\par 
This analysis concludes the investigation of constrained complex generalized supersymmetries
for maximal Clifford algebras. The extension of these results to the case of non-maximal Clifford algebras is encoded in the tables presented in the previous section and, with their knowledge, it can be easily computed.\par
An example of application of a constrained generalized supersymmetry was given in
\cite{lt3}, where the analytical continuation of the $M$-algebra to the $11$-dimensional Euclidean space was made possible by the introduction of a holomorphic complex generalized
supersymmetry. 

\section{Note on the construction of complex superparticles with tensorial central charges.}

We will discuss here the issues left opened in Section {\bf 5} concerning the construction
of the superparticle models with the bosonic tensorial central charges and complex spinors.
We will present the set of different choices for the metric ${\cal C}$ entering (\ref{Psquared}),
used to raise and lower spinorial indices.\par
The metric ${\cal C}$ has to be of the same form as ${\bf P}$ (see (\ref{Pmatrix})) entering the action (\ref{complexaction}),
with an upper-left (anti)symmetric block and an upper-right (anti)hermitian block. More specifically, ${\cal C}$ should be presented as in formula (\ref{Umatrix}), in terms of two 
(an (anti)symmetric and an (anti)hermitian) scalar matrices respectively denoted as $U$ and $V$.
Since $U$ and $V$ are both scalars, their available choices are therefore given by
$U\equiv {\widetilde C}$, $V\equiv {\widetilde A}$, where ${\widetilde C}$ denotes either the charge-conjugation matrix $C$ or, in the case of non-maximal Clifford algebras, one of the
products of $C$ with one or two external Gamma matrices, see the tables in Section {\bf 9}.
Similarly, ${\widetilde A}$ denotes either the matrix $A$ introduced in Section ${\bf 3}$
or, for non-maximal Clifford algebras, one of the products of $A$ with one or two external Gamma matrices.\par
It is convenient to denote with $\epsilon,\delta =\pm1$ (${\widetilde C}^T = \epsilon {\widetilde C}$, ${\widetilde A}^\dagger = \delta {\widetilde A}$)  the (anti)symmetry and (anti)hermitian properties of ${\widetilde C}$, ${\widetilde A}$ respectively.\par
Without loss of generality, three possible choices for ${\cal C}$ are at disposal. They are given by\par
{\em i})
\begin{eqnarray}
{\cal C} &=&\left(
\begin{tabular}{cc}
${\widetilde C}$ & $0$\\
$0$ &${\widetilde C}^\ast$
\end{tabular}
\right),
\end{eqnarray}
in this case ${\cal C}$ is (anti)symmetric in accordance with the sign of $\epsilon$;\par
{\em ii})
\begin{eqnarray}\label{iicasexi}
{\cal C} &=&\left(
\begin{tabular}{cc}
${0}$ & ${\widetilde A}$\\
$\xi {\widetilde A}^\ast$ &$0$
\end{tabular}
\right),
\end{eqnarray}
where $\xi$ is an arbitrary sign ($\xi=\pm 1$); in this case the (anti)symmetry property of
${\cal C}$ is specified by the sign of $\delta\xi$;
\par
{\em iii})
\begin{eqnarray}
{\cal C} &=&\left(
\begin{tabular}{cc}
${\widetilde C}$ & ${\widetilde A}$\\
$\epsilon\delta{\widetilde A}^\ast$ &${\widetilde C}^\ast$
\end{tabular}
\right),
\end{eqnarray}
the (anti)symmetry property of ${\cal C}$ is specified by the sign of $\epsilon$.
It should be noticed that in this last case an (anti)symmetric matrix ${\bf P}^2$ (
${\bf P}^2 ={\bf P}{\cal C}{\bf P}$) is only possible, for both non-vanishing
${\cal P}$, ${\cal R}$ entering ${\bf P}$, if the condition
\begin{eqnarray}
\epsilon&=&\delta
\end{eqnarray}
is matched.\par
The above three sets of choices for ${\cal C}$ completely specify the available actions for
the superparticles with tensorial central charges and complex spinors.\par
It is worth mentioning that, in the case of a space-time supporting Weyl spinors, both ${\widetilde C}$
and ${\widetilde A}$ are decomposed in $2\times 2$ block matrices. In presence of Weyl spinors
the metric ${\cal C}$ is not constructed with ${\widetilde C}$, ${\widetilde A}$ themselves,
but with their upper-left block projections $P({\widetilde C})$, $P({\widetilde A})$, see
formula (\ref{pweyl}).\par
The equations of motion of our class of models can be easily derived from the action (\ref{complexaction}). For our purposes it is not needed to write them explicitly. It is convenient, however, to present the constraints arising from the variations
$\delta e$, $\delta f$ of the lagrange multipliers entering (\ref{complexaction}).
Such constraints will be denoted with the symbols ``$X$"and ``$Y$", respectively.
In correspondence with the three above choices for ${\cal C}$ we get the following constraints
\\
{\em i})
\begin{eqnarray}
&&
X= {\cal P} {\widetilde C} {\cal P} + {\cal R}{\widetilde C}^\ast {\cal R}^\ast=0,\nonumber\\
&&Y = {\cal P} {\widetilde C} {\cal R} + {\cal R}{\widetilde C}^\ast {\cal P}^\ast=0;
\end{eqnarray}
{\em ii})
\begin{eqnarray}
&&
X= \xi {\cal R} {\widetilde A}^\ast {\cal P} + {\cal P}{\widetilde A} {\cal R}^\ast=0,\nonumber\\
&&Y = \xi {\cal R} {\widetilde A}^\ast {\cal R} + {\cal P}{\widetilde A} {\cal P}^\ast=0;
\end{eqnarray}
{\em iii})
\begin{eqnarray}
&&
X= {\cal P} {\widetilde C} {\cal P} + \epsilon\delta{\cal R}{\widetilde A}^\ast {\cal P}+ {\cal P}{\widetilde A}{\cal R}^\ast+ {\cal R}{\widetilde C}^\ast{\cal R}^\ast=0,\nonumber\\
&&Y = {\cal P} {\widetilde C} {\cal R} + \epsilon\delta{\cal R}{\widetilde A}^\ast {\cal R}+{\cal P}{\widetilde A}{\cal P}^\ast +{\cal R}{\widetilde C}^\ast{\cal P}^\ast=0.
\end{eqnarray}
In the next section we will 
compute, for the various different cases, the number of independent constraints. They depend
on the (anti)symmetry and (anti)hermitian properties of the lagrange multipliers $e$ and $f$
respectively, as well as the possible reality or imaginary conditions imposed on them. It is
useful to present here the table with the given number of constraints in association with 
$n$-component complex spinors. We have
{ {{\begin{eqnarray}&\label{counting}
\begin{tabular}{|c|c|c|c|c|}\hline
 $$&$sym.$&$antisym.$&$her.$&$antiher.$
 \\
 \hline
  $full$&$n^2+n$&$n^2-n$&$n^2$&$n^2$\\ \hline
  
  $real$&$\frac{1}{2}(n^2+n)$&$\frac{1}{2}(n^2-n)$&$\frac{1}{2}(n^2+n)$&$\frac{1}{2}(n^2-n)$\\ \hline
  
   $imag.$&$\frac{1}{2}(n^2+n)$&$\frac{1}{2}(n^2-n)$&$\frac{1}{2}(n^2-n)$&$\frac{1}{2}(n^2+n)$\\
  
  \hline

 \end{tabular}&\nonumber\\
&&\end{eqnarray}}} }  
}

\section{Constrained complex superparticles with tensorial central charges.}

We present now an application 
of the classification of the (constrained) complex generalized supersymmetries to the 
construction of the 
superparticle models with bosonic
tensorial central charges.\par
Let us recall at first the previous section results concerning the three available choices for the metric ${\cal C}$ entering
the action (\ref{complexaction}) (these three possibilities are denoted as {\em i}, {\em ii} and {\em iii}).\par
The bosonic sector of the theories under consideration is given by the constrained 
(see table (\ref{constrained})) or unconstrained tensorial
central charges entering the (\ref{Mhol}) and (\ref{Mher}) superalgebras.
Since the models depend on other, not purely algebraic, data, it is necessary 
to verify if some 
properties valid for the underlying algebra
are indeed applicable to the associated dynamical system. 
In particular the constraints need to be compatibile w.r.t. the equations of motion.
Similarly, the duality relations discussed in Section {\bf 10} between different formulations of the constrained supersymmetries, need a careful investigation in order to be promoted
as dualities between different formulations of the same theory. \par
It can be easily proven that all type of constraints (at least for some of the available choices of
${\cal C}$ entering the action (\ref{complexaction})) are indeed compatible
with the equations of motion, the only exceptions being the constraints labeled by $II$ and $III$ in Section {\bf 10}.
It is clear that these constraints are ``more difficult" to implement, due to the mixed requirements on ${\cal P}$ and ${\cal R}$, see (\ref{constrained})\footnote{It is worth stressing that their incompatibility with the dynamical systems under consideration does not dismiss them as possible viable constraints for
some other dynamical setting, $II$ and $III$ being, as pointed out in Section {\bf 10},  algebraically consistent.}.  \par
In this section we present some results concerning the bosonic sector of the tensorial superparticle models.
For each one of the dynamically compatible constraints $I$, $IV$, $V$, $VI$ and $VII$ 
of (\ref{dualities}) ($IV$, $V$ and $VI$ are analyzed for both their dual presentations) we compute the number of conditions on ${\bf P}^2$ (see (\ref{Psquared}))
given by the variation of the lagrange multipliers entering the (\ref{complexaction})) action.\par
This information tells us which are the admissible choices for the matrix ${\cal C}$ ({\em i}, {\em ii} or {\em iii}), since the number of lagrange multipliers constraints should not exceed
the number of bosonic degrees of freedom entering ${\cal P}$ and ${\cal R}$. \par
Our results are presented in terms of $n$, where $n$ is the number of components (complex counting) of the complex spinors entering the model. $n$ depends of course on the chosen space-time.
In a generic situation the inequality specifying that the number of lagrange multipliers are less or equal the number of bosonic degrees of freedom is valid for any value of $n$. In
a non-generic case such inequality is valid for at most some lowest values of $n$. An example
is given for the $IV$ ($a4$), see (\ref{constrained}), case with the ${iii}$ choice of ${\cal C}$ and 
$\epsilon = -1$, the number of bosonic degrees of freedom is $n^2+n$, while the number of
lagrange multipliers condition is given by $2n^2-n$. In this paper we will not investigate in detail the compatibility conditions associated to these non-generic cases.\par
The results depend on the signs $\epsilon$, $\delta$ denoting respectively the (anti)symmetric properties of ${\widetilde C}$ and
the (anti)hermitian properties of ${\widetilde A}$ (see the previous section). The ${ii}$ case further allows an arbitrary sign $\xi$ entering (\ref{iicasexi}). 
Most of the results obtained can be very conveniently summarized in terms of the (anti)symmetry properties of ${\bf P}^2$ (we recall that the symmetry of ${\cal C}$, tantamount to
the symmetry of ${\bf P}^2$, allows the introduction of a mass term).\par
It is worth presenting in some detail the analysis of the first few constraints. 
The remaining ones are analyzed along the same lines.
Let us denote with ``$\sharp$" the number of lagrange multiplier conditions.
We have at first for\\
$I$ (unconstrained supersymmetry)\\
that the number of bosonic degrees of freedom is $2n^2+n$.\\
All choices for ${\cal C}$ are acceptable, with
\\
~\\
\begin{tabular}{ll}{\em i})&$\sharp = 2n^2+n $ for $\epsilon = 1$,\\
 &$\sharp = 2n^2-n $ for $\epsilon = -1$;
 \\
 &\\
{\em ii})&$\sharp = 2n^2+n $ for $\xi\delta = 1$,\\
 &$\sharp = 2n^2-n $ for $\xi \delta = -1$;
 \\&\\{\em iii})&$\sharp = 2n^2+n $ for $\epsilon= \delta = 1$,\\
 &$\sharp = 2n^2-n $ for $\epsilon =\delta= -1$;
 \end{tabular}\\
 
 The choice $ii$ can be used, due to the arbitrarity of $\xi$, to impose, e.g., $2n^2+n$
 lagrange multiplier conditions even for spacetimes not supporting a
 symmetric ($\epsilon =1 $)
 ${\widetilde C}$ matrix.
 It should be further noticed that in all three cases $\sharp$ is the same according to
 the (anti)-symmetry properties of ${\bf P}^2$. \par
 The $IV$ ($a4$), see (\ref{constrained}), constraint gives us\\
that the number of bosonic degrees of freedom is $n^2+n$.\\
The acceptable choices for ${\cal C}$ are given by:
\\
~\\
\begin{tabular}{ll}{\em i})& (lagrange multipliers $f\equiv 0$)\\
&$\sharp = n^2+n $ for $\epsilon = 1$,\\
 &$\sharp = n^2-n $ for $\epsilon = -1$;
 \\
 &\\
{\em ii})&(lagrange multipliers $e\equiv 0$)\\
&$\sharp = n^2 $;
 \\&\\{\em iii})& not allowed for $\epsilon= 1$,\\
 &not generic  for $\epsilon =-1$;
 \end{tabular}\\
 
It is clear that the {\em ii} choice for the metric leads to an inequivalent theory w.r.t. the 
{\em i} choice, as it appears from the different counting of the lagrange multipliers conditions.
The above results for the $IV$ ($a4$) case can be summarized as follows. We get $\sharp= n^2+n, n^2$ conditions on
a symmetric ${\bf P}^2$ and $\sharp = n^2, n^2-n$ conditions on an antisymmetric ${\bf P}^2$.\par 
The $IV$ $(a4)$ constraint is dual to the $IV$ ($b2$) case. In this latter case
all three choices of the metric ${\cal C}$ are admissible (in the $iii$ case with the restriction that $\epsilon=\delta=1$). Due to the reality or imaginary requirements on the lagrange multipliers $e$, $f$ entering (\ref{complexaction}) we can verify that, for all three cases,
$\sharp= n^2+n, n^2$ for 
a symmetric ${\bf P}^2$ and $\sharp = n^2, n^2-n$ for an antisymmetric ${\bf P}^2$.\par
This is a strong indication that the algebraic duality described in Section {\bf 11}
promotes a duality at the level of the description of the constrained tensorial superparticle dynamics.\par
The same type of analysis can be repeated for the dual $V$ ($b3$) and $V$ ($c1$) constraints. 
In the $V$ ($b3$) case all three choices for the metric ${\cal C}$ are acceptable (in the
$iii$ case with the condition $\epsilon =\delta = 1$). In the $V$ ($c1$) constraint only
the $i$ and $ii$ choices for ${\cal C}$ are allowed. For both the dual $V$ constraints we obtain that
the number of lagrange multiplier conditions are given by $\sharp= n^2$ for a symmetric
${\bf P}^2$ and $\sharp =n^2,n^2-n$ for an antisymmetric ${\bf P}^2$.\par
Similarly, the $VI$ ($b4$) and ($c2$) constraints present the same type of description.
In both cases $i$ is admissible only for $\epsilon = -1$ (i.e. an antisymmetric ${\bf P}^2$),
with an associated $\sharp = \frac{1}{2}(n^2-n)$,
while $ii$ is always admissible and, according to the reality or imaginary condition on the lagrange multiplier $f$, we have $\sharp = \frac{1}{2}(n^2\pm n)$.\par
Finally, the $VII$ case gives us an acceptable $i$ choice for the ${\cal C}$ metric for $\epsilon=-1$, and an acceptable (with no condition) $ii$ choice. The number of lagrange multiplier conditions is given by $\sharp=\frac{1}{2}(n^2-n)$.\par
Some of the information here discussed can be very conveniently summarized in the following table, specifying the number of lagrange multiplier conditions $\sharp$ associated with 
a symmetric or an antisymmetric ${\bf P}^2$. We have
{{
{ {{\begin{eqnarray}&
\begin{tabular}{|l|l|l|}\hline \label{conditions}
 $cases$&$symm. \quad {\bf P}^2$&$antisym. \quad {\bf P}^2$\\ \hline 
$I$&$\sharp= 2n^2+n$&$\sharp = 2n^2-n$\\ \hline 
$IV$&$\sharp= n^2+n,n^2$&$\sharp= n^2,n^2-n$\\ \hline 
$V$&$\sharp= n^2$&$\sharp= n^2,n^2-n$\\ \hline 
$VI$&$\sharp= \frac{1}{2}(n^2\pm n)$&$\sharp= \frac{1}{2}(n^2\pm n)$\\ \hline 
$VII$&$\sharp= \frac{1}{2}(n^2-n)$&$\sharp= \frac{1}{2}(n^2-n)$\\ \hline
 \end{tabular}&\\ \nonumber
&&\end{eqnarray}}} }  
}}
For model building purposes it is also convenient to present a table with the allowed 
choices of the metric ${\cal C}$ in correspondence with each one of the dynamically
compatible constraint. We obtain (please notice that only {\em generic} choices for
${\cal C}$, in the sense specified above, have been inserted in the table)  
{{
{ {{\begin{eqnarray}&
\begin{tabular}{|l|l|l|l|}\hline \label{conditions2}
 $$&$i$&$ii$&$iii$\\ \hline 
$I$&$yes$&$yes$&$yes$\\ \hline 
$IV\,(a4)$&$yes$&$yes$&$no$\\ \hline
$IV\,(b2)$&$yes$&$yes$&$yes^\ast\, (\epsilon =1)$\\ \hline  
$V\,(b3)$&$yes$&$yes$&$yes^\ast\, (\epsilon =1)$\\ \hline 
$V\,(c1)$&$yes$&$yes$&$no$\\ \hline
$VI\,(b4)$&$yes^\ast\, (\epsilon = -1)$&$yes$&$no$\\ \hline
$VI\,(c2)$&$yes^\ast\, (\epsilon = -1)$&$yes$&$no$\\ \hline
$VII$&$yes^\ast\, (\epsilon = -1)$&$yes$&$no$\\ \hline
 \end{tabular}&\\ \nonumber
&&\end{eqnarray}}} }  
}}
The ``$\ast$" denotes which choices are consistent only for a specific value of $\epsilon$.\par

\section{Conclusions.}

The content of the present paper can be summarized as follows.
We made a detailed analysis of the real and complex generalized
supersymmetries and presented a classification, given by table (\ref{constrained}), of the
consistent constraints on the supersymmetry algebra. We proved that four of the seven classes of
constrained generalized supersymmetries admit a dual formulation. The bosonic sectors,
for the real cases and for each constrained complex case, were computed.\par
We constructed the generalized supersymmetries for each given space-time
in terms of the generalized supersymmetries in their associated {\em oxidized} spacetime. We recall that an oxidized spacetime corresponds to a maximal Clifford algebra (the Clifford algebras of the remaining spacetimes are a subset of the maximal Clifford algebras).
Generalized supersymmetries for non-maximal spacetimes can therefore be recovered through
a dimensional reduction from the oxidized generalized supersymmetries. Several important quantities, like the spinorial metric ${\widetilde C}$, ${\widetilde A}$ introduced in Section {\bf 11}, are easily computed in this unifying framework.
The construction of the non-maximal Clifford algebras and their supersymmetries was given in Section {\bf 7} for the real case. In Section {\bf 9} the derivation of the complex and quaternionic cases in terms of the quaternionic maximal Clifford algebras was presented.
\par
The obtained results were applied to the construction of the superparticles with tensorial central charges. The main ingredients in the construction of these models were presented in Section {\bf 5}. The subtleties and variants
in the construction of the complex superparticle models, depending on the existence of three distinct choices for the spinorial metric entering the action (\ref{complexaction}), were discussed in Section {\bf 12}. The constrained models were investigated in Section {\bf 13}. The compatibility of
the constrained generalized supersymmetries with respect to the equations of motion for each one of the three choices of the spinorial metric was checked.
The list of results was presented.
Concerning the algebraic dualities between constrained supersymmetries, at least for
some of the available constructions (i.e. for some of the admissible choices of the spinorial metrics) our results point towards a dual description of the constrained superparticle
theories. The matching in the counting of bosonic degrees of freedom and lagrange multiplier
conditions fulfills a necessary condition for the existence of such dynamical dualities.
It is outside the scope of the present paper and requires a detailed investigation of 
each given constrained tensorial superparticle model to check whether its dual formulations indeed reproduce an equivalent theory. 
\par
It is quite tempting to apply the present formulation and classification of constrained supersymmetries to several classes of dynamical systems. It looks promising, e.g., to investigate the possibility of a constrained twistorial formulation of the superparticles with
tensorial central charges (the twistorial formulation, see \cite{bls}, leads to a tower of massless higher spin particles). Some of our results are likely to be applicable to the systematic construction of the superstrings with tensorial central charges (whose physical
implications have been discussed in \cite{zlu}). Models defined in terms of superalgebras, like the supersymmetric extensions of the Chern-Simon supergravities in higher dimensions,
see \cite{htz}, are also natural candidates to be investigated with the present methods.\par
Finally, we mention that an immediate extension of the present results concerns the refinement of the classification of constrained generalized supersymmetries based on quaternionic division
algebras and quaternionic spinors, derived at first in \cite{top}.
$~$
\\$~$
\par
 {\large{\bf Acknowledgments.}} ~\\~\par

We are indebted to J. Lukierski for motivations and useful discussions.


\begin{thebibliography}{9}
\bibitem{rs} E. Sezgin and I. Rudychev, ``Superparticles, $p$-form coordinates and the BPS
condition", hep-th/9711128.
\bibitem{bl} I. Bandos and J. Lukierski, Mod. Phys. Lett. {\bf A 14} (1999) 1257.
\bibitem{bls} I. Bandos, J. Lukierski and D. Sorokin, Phys. Rev. {\bf D 61} (2000)
045002.
\bibitem{crt1} H.L. Carrion, M. Rojas and F. Toppan,
JHEP04 (2003) 040.
\bibitem{top} F. Toppan, JHEP09 (2004) 016.
\bibitem{abs} M.F. Atiyah, R. Bott and A. Shapiro, Topology
(Suppl. 1) {\bf 3} (1964) 3.
\bibitem{por} I.R. Porteous, ``Clifford Algebras and the Classical Groups", 
Cambridge Un. Press, 1995.
\bibitem{oku} S.Okubo, J. Math. Phys. {\bf 32} (1991) 1657; {\em ibid.}
{\bf 32} (1991) 1669.
\bibitem{dflv} R. D'Auria, S. Ferrara, M.A. Lledo and V.S. Varadarajan,
J. Geom. Phys. {\bf 40} (2001) 101 ; R. D'Auria, S. Ferrara and
M.A. Lledo, Lett. Math. Phys. {\bf 57} (2001) 123; S. Ferrara and
M.A. Lledo, Rev. Math. Phys. {\bf 14} (2002) 519.
\bibitem{acdp} D.V. Alekseevsky, V. Cort\'es, C. Devchand and A. van Proeyen,
``Polyvector Super-Poincar\'e Algebras", hep-th/0311107.
\bibitem{lt3} J. Lukierski and F. Toppan, Phys. Lett. {\bf B 584}
(2004) 315.
\bibitem{agit} J.A. de Azcarraga, J.P. Gauntlett, J.M. Izquierdo and P.K. Townsend,
Phys. Rev. Lett. {\bf 63} (1989) 2443.
\bibitem{tow} P. Townsend, hep-th/9712004, Cargese Lectures, 1997.
\bibitem{hls} R. Haag, J. \L opusza\'{n}ski and M. Sohnius, Nucl.Phys.
{\bf B 88} (1975) 257.
\bibitem{daf} R. D'Auria and P. Fr\'e, Nucl. Phys. {\bf B 201} (1982), 101; erratum
{\em ibid.} {\bf B 206} (1982) 496.
\bibitem{gs} R. G\"uven, Phys. Lett. {\bf B 276} (1992) 49.
\bibitem{ste} K.S. Stelle, ``Lectures on Supergravity p-branes", 1996 ICTP Summer School,
hep-th/9701088.
\bibitem{west} P. West, JHEP 0008 (2000) 007.
\bibitem{htz} M. Hassaine, R. Troncoso and J. Zanelli, Phys. Lett. {\bf B 596} (2004) 132.
\bibitem{fro} C. Fronsdal, ``Massless Particles, Ortosymplectic Symmetry and Another Type of Kaluza-Klein Theory",
in ``Essays on Supersymmetry", Reidel (1986) (Math. Phys. Studies, v. 8).
\bibitem{sor} D. Sorokin, ``Introduction to the Classical Theory of Higher Spin", hep-th/0405069.
\bibitem{zlu} A.A. Zheltukhin and U. Lindstr\"om, Nucl. Phys. Proc. Suppl. {\bf 102}
(2002) 126; JHEP 0201 (2002) 034; A.A. Zheltukhin and D. V. Uvarov, JHEP 0208 (2002) 008;
Phys. Lett. {\bf B 565} (2003) 229.
\bibitem{kt} T. Kugo and P. Townsend, Nucl. Phys. {\bf B 221}
(1983) 357.
\bibitem{lpss} H. Lu, C.N. Pope, E. Sezgin and K.S. Stelle, Nucl. Phys. {\bf B 456} 
(1995) 669; K.S. Stelle, ``Revising Supergravity and Super Yang-Mills
Renormalization" in ``{New Developments in Fundamental Interaction Theories}",
AIP 2001, eds. J. Lukierski and J. Rembieli\'nski, p. 108.
\bibitem{brsc} L. Brink and J. Schwarz, Phys. Lett. {\bf B 100} (1981) 310.
\bibitem{lt1} J. Lukierski and F. Toppan, Phys. Lett. {\bf B 539},
(2002) 266.

\end{thebibliography}
\end{document}